\documentclass[aps,physrev,preprint,superscriptaddress]{revtex4-2}

\usepackage{amssymb,amsthm}
\usepackage{braket}
\usepackage{color}
\usepackage{booktabs,longtable}
\usepackage{mathtools}
\usepackage{tikz-cd}
\usepackage{hyperref}
\usepackage{cleveref}

\newcommand{\abs}[1]{\left\vert#1\right\vert}
\newcommand{\defeq}{\vcentcolon=}
\DeclareMathOperator{\EX}{\mathbb{E}}
\DeclareMathOperator{\Tr}{Tr}
\newcommand{\diag}{\mathop{\mathrm{diag}}}

\newtheorem*{theorem}{Theorem}
\theoremstyle{remark}
\newtheorem*{remark}{Remark}

\begin{document}


\title{Oracle problems as communication tasks and optimization of quantum algorithms}


\author{Amit Te'eni}
\email[Corresponding author, ]{amit.teeni@biu.ac.il}
\affiliation{Faculty of Engineering and Institute of Nanotechnology and Advanced Materials, Bar Ilan University, Ramat Gan, Israel}

\author{Zohar Schwartzman-Nowik}
\affiliation{Faculty of Engineering and Institute of Nanotechnology and Advanced Materials, Bar Ilan University, Ramat Gan, Israel}
\affiliation{Department of Computer Science and Engineering, Hebrew University of Jerusalem, Jerusalem, Israel}

\author{Marcin Nowakowski}
\affiliation{Faculty of Applied Physics and Mathematics, Gdansk University of Technology, 80-952 Gdansk, Poland}
\affiliation{National Quantum Information Center, 80-309 Gdansk, Poland}
\affiliation{ObserveQ, 80-890 Gdansk, Poland}

\author{Pawe{\l} Horodecki}
\affiliation{International Centre for Theory of Quantum Technologies (ICTQT), University of Gdansk, 80-309 Gdansk, Poland}
\affiliation{Faculty of Applied Physics and Mathematics, Gdansk University of Technology, 80-952 Gdansk, Poland}

\author{Eliahu Cohen}
\affiliation{Faculty of Engineering and Institute of Nanotechnology and Advanced Materials, Bar Ilan University, Ramat Gan, Israel}
\affiliation{Institute for Quantum Studies, Chapman University, Orange, California 92866, USA}



\begin{abstract}
Quantum query complexity studies the number of queries needed to learn some property of a black box. A closely related question is how well an algorithm can succeed with this learning task using only a fixed number of queries. In this work, we propose measuring an algorithm's performance using the mutual information between the output and the actual value. The task of optimizing this mutual information using a single query, is similar to a basic task of quantum communication, where one attempts to maximize the mutual information of the sender and receiver. We make this analogy precise by splitting the algorithm between two agents, obtaining a communication protocol. The oracle's target property plays the role of a message that Alice encodes into a quantum state, which is subsequently sent over to Bob. The first part of the algorithm performs this encoding, and the second part measures the state and aims to deduce the message from the outcome.
Moreover, we formally consider the oracle as a separate subsystem, whose state records the unknown oracle identity. Within this construction, Bob's optimal measurement basis minimizes the quantum correlations between the two subsystems.
We also find a lower bound on the mutual information, which is related to quantum coherence. These results extend to multiple-query non-adaptive algorithms. As a result, we describe the optimal non-adaptive algorithm that uses at most a fixed number of queries, for any oracle classification problem.
Crucially, this mutual-information perspective carries direct practical utility for algorithmic design, providing the theoretical foundation to optimize iterative subroutines in hybrid quantum--classical schemes (as recently demonstrated for Quantum Likelihood Estimation in~[Alon Levi et al 2026 Quantum Sci. Technol. 11 015029]). Within the present work, we apply this framework to analyze the stage-by-stage information flow and track partial progress in several standard quantum algorithms, including the Deutsch--Jozsa, Bernstein--Vazirani, and Shor--Kitaev algorithms.
\end{abstract}


\maketitle

\section{Introduction}
Quantum query complexity explores the performance of quantum algorithms in terms of the number of queries they make to an oracle~\cite{buhrman2002complexity}. There are at least two good reasons to study query complexity. First, many important quantum algorithms solve oracle problems, including some of the first examples of quantum advantage (such as the Deutsch--Jozsa~\cite{Deutsch1992}, Bernstein--Vazirani~\cite{Bernstein1997} and Simon~\cite{simon1997power} algorithms). Second, query complexity is much simpler to study than time complexity (the number of gates used in an algorithm). Indeed, it is sometimes possible to prove lower bounds on query complexity~\cite{bennett1997strengths,beals2001quantum,aaronson2004quantum,Reichardt2011Reflections,zhang2023quantum,lombardi2023one} (lower bounds are usually difficult to obtain in both classical and quantum complexity theory). On the other hand, results about query complexity may allow us to obtain insight regarding time complexity.
Therefore, quantum query complexity remains widely studied, with many interesting problems still open \cite{aaronson2021open,aaronson2022much}.

One avenue of research seeks to study the origins of \textit{quantum speedup}, i.e. the computational advantage of quantum computers with respect to classical ones. The role played by various resources in expediting quantum computation has been studied extensively \cite{Matera2016,Howard2014,Vidal2003,Hillery2016,Stahlke2014,Ma2016,Cai2015,Datta2008,Horodecki2009,Nest2013,Biham2004,Linden2001,Ekert1998,Jozsa2003,Anand2016,Vedral2010,Lloyd1999,Jozsa1997}, and remains an active area of research. One of the quantum resources that seems to be essential for quantum speedup is quantum discord~\cite{Datta2008,Vedral2010,Brodutch2013,Cable2015,Modi2012}. However, research in this direction has focused on mixed-state quantum computation, since for pure states the notion of quantum discord coincides with entanglement entropy. Interestingly, quantum advantage in communication tasks seems to require similar resources. Indeed, quantum nonlocality is known to be a resource for quantum communication~\cite{buhrman2016quantum}, and discord in particular~\cite{madhok2013quantum}. Relations between quantum computation and communication were studied as well~\cite{buhrman1998quantum,de2001quantum,de2002quantum,brassard2003quantum,spiller2006quantum,buhrman2010nonlocality,hoza2017quantum,chakraborty2019quantum,chakraborty2020role,blikstad2022nearly,chakraborty2025scalable,rout2025unbounded}.

In this work, we develop a framework for studying single-query quantum algorithms solving oracle classification tasks. We fix a probability distribution over the set of oracles that may appear in the black box. The oracle class---the correct solution to the classification problem---thus becomes a random variable. We measure the algorithm's performance via the mutual information between this solution and the algorithm's output. This perspective eliminates the need to consider any classical post-processing performed after measuring the quantum state. We derive an upper bound on the mutual information via an analogy between our setup and quantum communication tasks.

Using our analogy, we prove a theorem that characterizes the optimal single-query algorithm. To do so, we consider the oracle as a separate subsystem in a classical mixture of all its possible states.
We also find a lower bound on the mutual information, derived from a measure of quantum coherence. Finally, we study how several physical and information-theoretic quantities of interest vary throughout known algorithms.

\subsection{Main contributions}
The main contribution of this paper is a theorem that describes the optimal quantum algorithm for an arbitrary oracle classification task. Here, optimality is defined with respect to the mutual information between the correct result and the algorithm's output. 
As finding the optimal algorithm is generally difficult, we focus in this work on non-adaptive algorithms with a fixed number of queries. 
The optimal algorithm is given by solving a double optimization problem, which involves information-theoretic quantities such as Holevo's quantity, mutual information and quantum discord, thereby highlighting their fundamental significance for quantum computation. Our results also demonstrate the utility of measuring algorithms' performance via mutual information.

While the primary focus of this paper is establishing the theoretical framework, we emphasize that this mutual-information perspective carries immediate practical utility for algorithmic design. Specifically, this framework has already been operationalized to significantly improve hybrid quantum--classical algorithms. In recent work~\cite{levi2025optimalquantumlikelihoodestimation}, we applied these exact principles to Quantum Likelihood Estimation (QLE) for Hamiltonian learning. By modeling each QLE iteration as a single-query oracle task and utilizing the optimality theorem developed here to maximize mutual information, we were able to systematically minimize quantum discord and substantially accelerate the algorithm's convergence. This demonstrates that reformulating oracle problems as communication tasks is not merely a conceptual exercise, but a concrete tool for optimizing NISQ-era algorithms.

\subsection{Related work}
Although independently derived, one may notice slight similarity between our approach and the one proposed by Buhrman, Cleve and Wigderson~\cite{buhrman1998quantum}, in the sense that both constructively demonstrate the existence of a protocol for a quantum communication task, given a quantum algorithm for an oracle problem. However, our analogy always yields a very simple communication task, where Alice wishes to communicate a single message to Bob.
Our analogy is used to find an upper bound on the mutual information obtainable in a single query. This is in contrast to existing works, where the total query complexity of the computational problem is related to the bit complexity of the communication task. Therefore, the analogy presented here yields insights regarding information-theoretic quantities, including quantum coherence and discord.

We note that the problem of finding optimal quantum query algorithms has been addressed from a different perspective by Belovs~\cite{belovs2015variationsquantumadversary}, who provided tight upper and lower bounds for adaptive query complexity (which encompasses the non-adaptive setting) by formulating the task as an exponentially large semidefinite program (SDP). While both the SDP approach and our framework ultimately lead to optimization problems that are generally difficult to solve for arbitrary tasks, their fundamental objectives and insights differ. Belovs's SDP is designed to minimize query complexity for a target success probability. In contrast, our approach fixes the query count and optimizes the extracted mutual information. Crucially, rather than yielding a purely mathematical program, our formulation analytically links computational optimality to physical, information-theoretic resources---specifically, the Holevo quantity, quantum discord, and relative entropy of coherence. This allows us not only to characterize the optimal algorithm but also to track the physical flow of information throughout the stages of the quantum circuit.

Finally, regarding our representation of the oracle as a physical subsystem, our classical mixture construction is intimately related to an older one by Ambainis~\cite{ambainis2000quantum} (see also \cite{hoyer2007negative}). In that work, the oracle subsystem is in a coherent superposition of its possible states; indeed, our classical mixture can be viewed as the result of applying a dephasing channel to that pure state (with the oracle identity traced out). However, working explicitly with this classical-quantum mixed state is conceptually advantageous for our purposes, as it naturally allows the (non-minimized) quantum discord between the computer and oracle subsystems to play a key role in our characterization theorem.

\subsection{Paper organization}
The remainder of the paper is organized as follows.
In \Cref{sec:setting} we define the formal setting, detailing the oracle classification problem and the scope of the quantum algorithms considered. \Cref{sec:framework} outlines our main results: we establish our general information-theoretic framework, derive the optimality conditions for single-query algorithms, prove upper and lower bounds on algorithmic performance, and demonstrate the direct application of these principles to optimizing hybrid quantum--classical schemes such as Quantum Likelihood Estimation.
In \Cref{sec:analysis} we apply this framework to analyze the stage-by-stage information flow in several standard quantum algorithms, including the Deutsch--Jozsa, Bernstein--Vazirani, and phase estimation algorithms, and sketch the analysis for the Shor--Kitaev algorithm, using Simon's algorithm as an example. Finally, in \Cref{sec:discussion} we summarize our findings and discuss possible directions for future work.

\section{\label{sec:setting}Preliminaries}
In this section, we establish the formal mathematical setting for our study by defining the oracle classification problem and delineating the structure of the quantum algorithms under consideration.

\subsection{\label{oracle_problems}Single-query oracle classification}
Here we introduce the setting for the type of computational problems we study, as well as the notation used later. Our definitions are fairly similar to the ones found in several papers~\cite{montanaro2007structure,montanaro2016quantum,chailloux2018note}. We do not claim that the setting presented here is the most general or important one possible; however, we do provide justification for some of our choices here.

An oracle classification problem is described by the following data:
\begin{itemize}
	\item A set $ \mathcal{F} $ of all allowed oracle ``identities'';
	
	\item A partition: $ \mathcal{F} = \bigsqcup_{j \in \mathcal{J}} A_j $, where $ \bigsqcup$ denotes a disjoint union and the sets $A_j$ are all nonempty;
	
	\item A protocol for oracle queries, in the form of unitary gates $ \left\{ U_f \mid f \in \mathcal{F} \right\} $.
\end{itemize}

Usually the problem is, given an unknown oracle $ f \in \mathcal{F} $, to find its \textit{class} with respect to the partition. In other words, it is required to find the unique index $j \in \mathcal{J}$ for which $ f \in A_j $, using as few oracle queries as possible. However, we consider a strongly relevant variation where the problem contains a fourth component:
\begin{itemize}
	\item A probability distribution $ p_f \defeq \Pr \left( F = f \right) $ on $ \mathcal{F} $, i.e. a random variable $F$ that obtains its values in $ \mathcal{F} $.
\end{itemize}
Introducing a prior distribution is a fundamentally important variation; by Yao's minimax principle~\cite{Yao1977Probabilistic}, the worst-case success probability of a quantum algorithm over all oracles is equal to its average success probability against an adversarially chosen distribution (see also \Cref{app:prob}). While such probabilistic settings typically aim to maximize this average success probability using a single oracle query, our objective here is to maximize the \emph{mutual information} between the true oracle class and the algorithm's output. As we demonstrate, optimizing this information-theoretic objective is intimately related to, yet distinct from, maximizing the probability of success.

Our guiding example would be the Deutsch--Jozsa-$k$ problem (hereon denoted $DJ(k)$), where:
\begin{itemize}
	\item $ \mathcal{F} $ is the set of all functions $ \left\{ 0,1 \right\}^k \rightarrow \left\{ 0,1 \right\} $ which are either constant or balanced;
	
	\item $ \mathcal{F} $ is partitioned as a disjoint union of the subsets $ A_b, A_c $ of balanced and constant functions, respectively;
	
	\item The protocol is given by the unitary gates $U_f$, defined on computational basis states as:
	\begin{equation}
		U_f \ket{x} \otimes \ket{a} = \ket{x} \otimes \ket{a \oplus f \left( x \right) } , \quad x \in \left\{ 0,1 \right\}^k , a \in \left\{ 0,1 \right\},
	\end{equation}
	where $ \oplus $ denotes addition modulo $2$.
\end{itemize}

As mentioned before, many problems (such as $DJ(k)$) are not readily equipped with a probability distribution. However, there are at least two ``natural'' choices: the uniform distribution over $ \mathcal{F} $, and the uniform distribution with respect to the partition. The latter is defined explicitly by:
\begin{equation}
	p_f = \frac{1}{ \abs{\mathcal{J}} \abs{A_j} } , \quad f \in A_j .
\end{equation}
This distribution ensures that $f$ belongs to each of the classes $A_j$ with equal probabilities, and that the distribution within each class is uniform. Let $J$ denote the random variable that keeps track of the class $A_j$; that is, $ J \defeq \tilde{j} \left( F \right) $ where $\tilde{j}$ is the function that assigns to each $f \in \mathcal{F}$ its corresponding class, i.e. the unique $j \in \mathcal{J}$ such that $f \in A_j$.

As an aside, note that in $DJ(k)$ we have $ \abs{\mathcal{J}}=2 $. Classification problems with this property are also known as oracle decision problems; and for such problems, one often considers $ \mathcal{F} $ as a subset of $ \left\{ 0,1 \right\}^N $ for some $N$ and identifies $ \mathcal{J} $ with $ \left\{ 0,1 \right\} $. Solving the oracle problem is then referred to as computing the partial Boolean function $\tilde{j} : \mathcal{F} \rightarrow \left\{ 0, 1 \right\} $. Single-query quantum algorithms for partial Boolean functions were studied extensively~\cite{jozsa1991characterizing,Aaronson_2016_Polynomials,PhysRevA.101.022325,xu2021partial,qiu2021exact,ye2023characterization}.

The polar opposite of decision problems is the \textit{oracle identification problem}, where the task is to identify the unknown oracle. Of course, this is a special case of oracle classification, given by the partition of $\mathcal{F}$ into singletons. Some instances of oracle identification are well-known, including unstructured search (solved by Grover's algorithm) and the Bernstein--Vazirani problem. Moreover, the query complexity of the general oracle identification problem was studied extensively \cite{ambainis2004quantum,ambainis2004robust,ambainis2007improved,kothari2013optimal,taghavi2022simplified}, with focus on generic algorithms that solve \textit{all} oracle identification problems parameterized by $N$ and the size of the set $\mathcal{F} \subseteq \left\{ 0,1 \right\}^N $.

Further remarks regarding the problem formulation appear in \Cref{app:formulation}.

\subsection{\label{algo}Single-query algorithms}
A general quantum algorithm for an oracle that uses only a single query has the following structure:
\begin{enumerate}
	\setcounter{enumi}{-1}
	\item Initialize the quantum computer to be in some initial $ n$-qubit state $ \ket{\psi_0} $, where $ n \geq m $; \label{init}
	
	\item Apply some unitary gate $ V $; \label{V}
	
	\item Perform a single oracle query $U_f \otimes I$, where $ U_f$ acts on the first $m$ qubits and the identity matrix acts on the remaining $n-m$ qubits; \label{query}
	
	\item Apply an additional unitary gate $W$; \label{W}
	
	\item Measure all or some of the $n$ qubits in the computational basis; obtain a bit-string $y$ (hereon referred to as the \textit{measurement outcome}); \label{measure}
	
	\item Based on the bit-string $y$, provide an output $\hat{j} \in \mathcal{J}$. This chosen output is given by some function $g : \left\{ 0,1 \right\}^n \rightarrow \mathcal{J}$, i.e. $ \hat{j} = g \left( y \right) $. \label{output}
\end{enumerate}
where the algorithm's performance is either the probability of finding the right class, i.e. $ p_s \defeq \Pr \left( \hat{j} = j \right) $; or the mutual information $ I \left( J; Y \right) $ (see remarks below). By $ I \left( J; Y \right) $ we mean the (Shannon) mutual information of the (classical) random variables $J$ and $Y$, where $Y$ is the random variable that obtains the values $y$ in step \ref{measure} (a more explicit definition is given at the beginning of \Cref{sec:framework}).

\paragraph*{Remarks:}
\begin{enumerate}
	\item The number $m$ is determined by the problem setting via the dimensionality of the oracle unitaries $U_f$. Using $n > m$ qubits effectively allows providing a mixed state as the input to the oracle, as well as implementing a general POVM instead of a projective measurement in step \ref{measure}. Hereon we mostly assume some fixed value of $n$ and use $U_f$ to mean $U_f \otimes I$, in a slight abuse of notation.
	
	\item For any fixed choice of an initial state and gates $V, W$, one may compute the conditional probability distribution $ \Pr \left( J = j \vert Y = y \right) $ (where $J$ is a random variable as well, whose distribution depends on the distribution we have chosen for $F$). It is clear that any algorithm that does not output in the final step the value of $\hat{j}$ with highest conditional probability (given $y$ from the fifth step) can be improved. Thus, once we have decided upon the first five steps, there is always an obvious choice for the final one.
	
	\item The algorithm succeeds with probability $1$, i.e. $p_s = 1$, if and only if $ I \left( J;Y \right) = H \left( J \right) $, where $ H \left( J \right) $ is the Shannon entropy of the random variable $J$.
	
	\item More generally, Fano's inequality provides a relation between the probability of success and the mutual information. Assuming that $J$ has the uniform distribution, the probability of success $p_s$ is bounded from above according to:
	\begin{equation}
		p_s \leq \frac{I \left( J; Y \right) +1}{\log \abs{\mathcal{J}}} .
	\end{equation}
	Thus, low mutual information implies low probability of success. Conversely, mutual information also provides a lower bound on the probability of success:
	\begin{equation}\label{reverse_Fano}
		p_s \geq 2^{ I \left( J; Y \right) - \log \abs{\mathcal{J}} } .
	\end{equation}
	
	\item In what follows, we focus on the mutual information $ I \left( J; Y \right) $ as our chosen measure of performance. Since step \ref{output} does not affect the mutual information, we shall mostly ignore it from now on. For further discussion and computations regarding the probability of success, including a proof of \eqref{reverse_Fano}, see \Cref{app:prob}.
\end{enumerate}

Let us emphasize an important point: the problem setting already determines step \ref{query} (up to choosing the total number of qubits $n$); step \ref{measure} is always the same if we measure all qubits, which we can do without losing generality (we may simply ignore some bits of the measured bit string); and for any choice of steps \ref{init}, \ref{V} and \ref{W}, there is an obvious choice for step \ref{output}. Moreover, we can always choose $ \ket{\psi_0} = \ket{0}^{\otimes n} $ and ``absorb'' the change in the initial state into $V$. In other words, the various possible ``reasonable'' quantum algorithms for a given single-query oracle problem essentially differ in the value of $n$ and their pre- and post-query gates, $V$ and $W$. In the next section we show how to choose the optimal post-query unitary $W$ as well, given fixed choices for $n$ and $V$.

As an example, let us describe a quantum algorithm for the $DJ \left(k\right)$ problem; this is the Deutsch--Jozsa algorithm, denoted $ QDJ \left(k\right) $.

\paragraph*{The Deutsch--Jozsa algorithm $QDJ \left( k \right)$}
\begin{enumerate}
	\item Initialize the $(k+1)$-qubit state $ \ket{\psi_0} = \ket{0}^k \otimes \ket{1} $
	
	\item Apply the unitary gate $V = H^{\otimes (k+1)}$
	
	\item Perform a single oracle query $U_f$
	
	\item Apply an additional unitary gate, $W = H^{\otimes k} \otimes \mathbb{I}$
	
	\item Measure the first $k$ qubits in the computational basis (equivalently, measure all of them but ignore the outcome of the final qubit)
	
	\item If the measurement result was $ 0 \dots 0 $ (i.e. the all-zeros bit string), the output is $\hat{j}=c$ (constant); otherwise, $\hat{j}=b$ (balanced).
\end{enumerate}
$ QDJ \left( k \right) $ is a deterministic quantum algorithm for the $ DJ \left( k \right) $ problem that always succeeds. It is known that this problem does not admit any single-query classical algorithms that always succeed, meaning it has a \textit{quantum advantage}.

\subsection{\label{non-adaptive}Non-adaptive algorithms}
A multiple-query algorithm for an oracle problem can be either adaptive or non-adaptive. Here, ``adaptive'' means that outcomes of queries can be used to compute the inputs for later queries. If an algorithm makes several oracle queries in a non-adaptive manner, then it can be converted to an algorithm that makes all its queries at the same time~\cite{Montanaro2010Nonadaptive}. Thus, we define a non-adaptive $t$-query algorithm as follows:
\begin{enumerate}
	\item Initialize the quantum computer to be in some initial $ n$-qubit state $ \ket{\psi_0} $, where $ n \geq m t $;
	
	\item Apply some unitary gate $ V $;
	
	\item Perform $t$ simultaneous oracle queries $U_f^{\otimes t} \otimes I$, where $ U_f^{\otimes t}$ acts on the first $m t$ qubits and the identity matrix acts on the remaining $n-mt$ qubits;
	
	\item Apply an additional unitary gate $W$;
	
	\item Measure all or some of the $n$ qubits in the computational basis; obtain a bit-string $y$ (hereon referred to as the \textit{measurement outcome});
	
	\item Based on the bit-string $y$, provide an output $\hat{j} \in \mathcal{J}$. This chosen output is given by some function $g : \left\{ 0,1 \right\}^n \rightarrow \mathcal{J}$, i.e. $ \hat{j} = g \left( y \right) $.
\end{enumerate}
Note that any $t$-query non-adaptive algorithm can be thought of as single-query algorithm with a different (``stronger'') oracle, with $m' = m t$ and $ U'_f = U_f^{\otimes t} $. Thus, we may treat it as a single-query algorithm, and by virtue of this observation the entire discussion would also apply to multiple-query non-adaptive algorithms.

\section{\label{sec:framework}Theoretical framework and optimization}
As stated before, the measurement outcome of a quantum algorithm is generally a random variable; it is determined by the outcome of measuring all or some of the qubits in the computational basis. Thus, $ Y $ is defined to be the outcome of measuring the final state of the computation with respect to the observable $ Z^{\otimes k}\otimes I_2^{\otimes \left(n-k\right)} $, where $n$ is the total number of qubits and $ k \leq n $ is the number of measured qubits. Equivalently, we may take $Y$ to be the outcome of measuring \textit{all} qubits of a suitable reduced subsystem, i.e. we first trace out the non-measured qubits and then measure all remaining qubits in the computational basis. In what follows we mostly use the latter point of view.

For the moment, let us disregard the oracle classes $A_j$ and assume that our task is finding the exact oracle identity $f$. This task can be reformulated as follows: an agent (the oracle) prepares the pure state $ \ket{\psi_\mathrm{fin} \left( f \right)} $ with probability $p_f$, and the other party is tasked with deducing the index $f$ from a measurement on the following mixed state:
\begin{equation}\label{rho_Y_def}
	\rho_Y \defeq \EX_{f \in \mathcal{F}} \left[ \ket{ \psi_\mathrm{fin} \left( f \right) } \bra{ \psi_\mathrm{fin} \left( f \right) } \right] = \sum_{f \in \mathcal{F}} p_f T_f \ket{ \psi_0 } \bra{ \psi_0 } T_f^\dagger ,
\end{equation}
where $ T_f = W U_f V $ is the total quantum circuit and $ \ket{\psi_\mathrm{fin} \left( f \right)} = T_f \ket{\psi_0} $ is the final state of the computation (before measurement).
Thus, the state \eqref{rho_Y_def} defines precisely the type of mixture for which Holevo's bound applies. The subscript $Y$ for $ \rho_Y $ is used to clarify that this state acts on a Hilbert space with a distinguished orthonormal basis (i.e. the computational basis), whose elements correspond to the values of the random variable $Y$. Thus $ \rho_Y $ defines the probability distribution of the random variable $Y$, $ \Pr \left( Y=y \right) \defeq \braket{y \vert \rho_Y \vert y} $.

Recall that Holevo's bound deals with the following communication task: Alice randomly generates a letter $x$ from some finite alphabet $ \mathcal{X} $, and then prepares a quantum state $ \rho_x $ corresponding to that letter. She then sends the state to Bob, so he obtains the mixture $ \rho = \sum_{x \in \mathcal{X}} p_x \rho_x $.
Bob measures $\rho$ in some basis and then uses the outcome to deduce the letter. Holevo's bound provides an upper bound on the mutual information between the random letter $X$ and the measurement outcome $Y$ (here $X,Y$ are the corresponding classical random variables):
\begin{equation}\label{Holevo_general}
	I \left( X; Y \right) \leq S \left( \rho \right) -\sum_x p_x S \left( \rho_x \right) .
\end{equation}
The RHS is called the \textit{Holevo quantity}, denoted $ \chi \left( \left\{ p_x, \rho_x \right\}_{x \in \mathcal{X}} \right) $.

In our setting, the unknown oracle identity $f$ (and later the property $j$) is the random letter; and the quantum circuit plays the role of Alice, preparing a state based on $f$. The post-query unitary $W$ chooses the measurement basis, and $Y$ is the outcome of this final measurement.
Thus we obtain an upper bound on the mutual information, $ I \left( F; Y \right) \leq S \left( \rho_Y \right) $. Note that if the first $k$ qubits are measured then we can redefine $ \rho_Y $ as the appropriate reduced state of \eqref{rho_Y_def}, and the inequality $ I \left( F; Y \right) \leq S \left( \rho_Y \right) $ still holds.

Now we may derive a more general result, corresponding to our actual task of finding the oracle class $A_j$. We rewrite $ \rho_Y $ as follows:
\begin{equation}\label{rho_Yj_def}
	\rho_Y = \sum_{j \in \mathcal{J}} p_j \sigma_j \; , \quad \sigma_j \defeq \frac{1}{p_j} \sum_{f \in A_j} p_f T_f \ket{ \psi_0 } \bra{ \psi_0 } T_f^\dagger ,
\end{equation}
where $ p_j \defeq \sum_{f \in A_j} p_f $. We still have two agents, where one wishes to communicate a letter of the alphabet $ \mathcal{J} $ to the other. The first agent prepares the mixed state $ \sigma_j $ with probability $ p_j $, and sends the mixture $ \rho_Y $ to Bob. Bob then measures $ \rho_Y $ in the computational basis, resulting in the bit string $Y$. Holevo's bound then provides an upper bound on the mutual information:
\begin{equation}
	I \left( J; Y \right) \leq \chi \left( \left\{ p_j, \sigma_j \right\}_{j \in \mathcal{J}} \right) = S \left( \rho_Y \right) -\sum_{j \in \mathcal{J}} p_j S \left( \sigma_j \right) .
\end{equation}
Hereon we write $ \chi $ for $ \chi \left( \left\{ p_j, \sigma_j \right\}_{j \in \mathcal{J}} \right) $. Again, we can replace all states $\sigma_j$ with their reduced states corresponding to the measured registers, and the inequality still holds.

\begin{figure}
	\def\svgwidth{\columnwidth}
\begingroup%
  \makeatletter%
  \providecommand\color[2][]{%
    \errmessage{(Inkscape) Color is used for the text in Inkscape, but the package 'color.sty' is not loaded}%
    \renewcommand\color[2][]{}%
  }%
  \providecommand\transparent[1]{%
    \errmessage{(Inkscape) Transparency is used (non-zero) for the text in Inkscape, but the package 'transparent.sty' is not loaded}%
    \renewcommand\transparent[1]{}%
  }%
  \providecommand\rotatebox[2]{#2}%
  \newcommand*\fsize{\dimexpr\f@size pt\relax}%
  \newcommand*\lineheight[1]{\fontsize{\fsize}{#1\fsize}\selectfont}%
  \ifx\svgwidth\undefined%
    \setlength{\unitlength}{737.00787402bp}%
    \ifx\svgscale\undefined%
      \relax%
    \else%
      \setlength{\unitlength}{\unitlength * \real{\svgscale}}%
    \fi%
  \else%
    \setlength{\unitlength}{\svgwidth}%
  \fi%
  \global\let\svgwidth\undefined%
  \global\let\svgscale\undefined%
  \makeatother%
  \begin{picture}(1,0.66153846)%
    \lineheight{1}%
    \setlength\tabcolsep{0pt}%
    \put(0,0){\includegraphics[width=\unitlength,page=1]{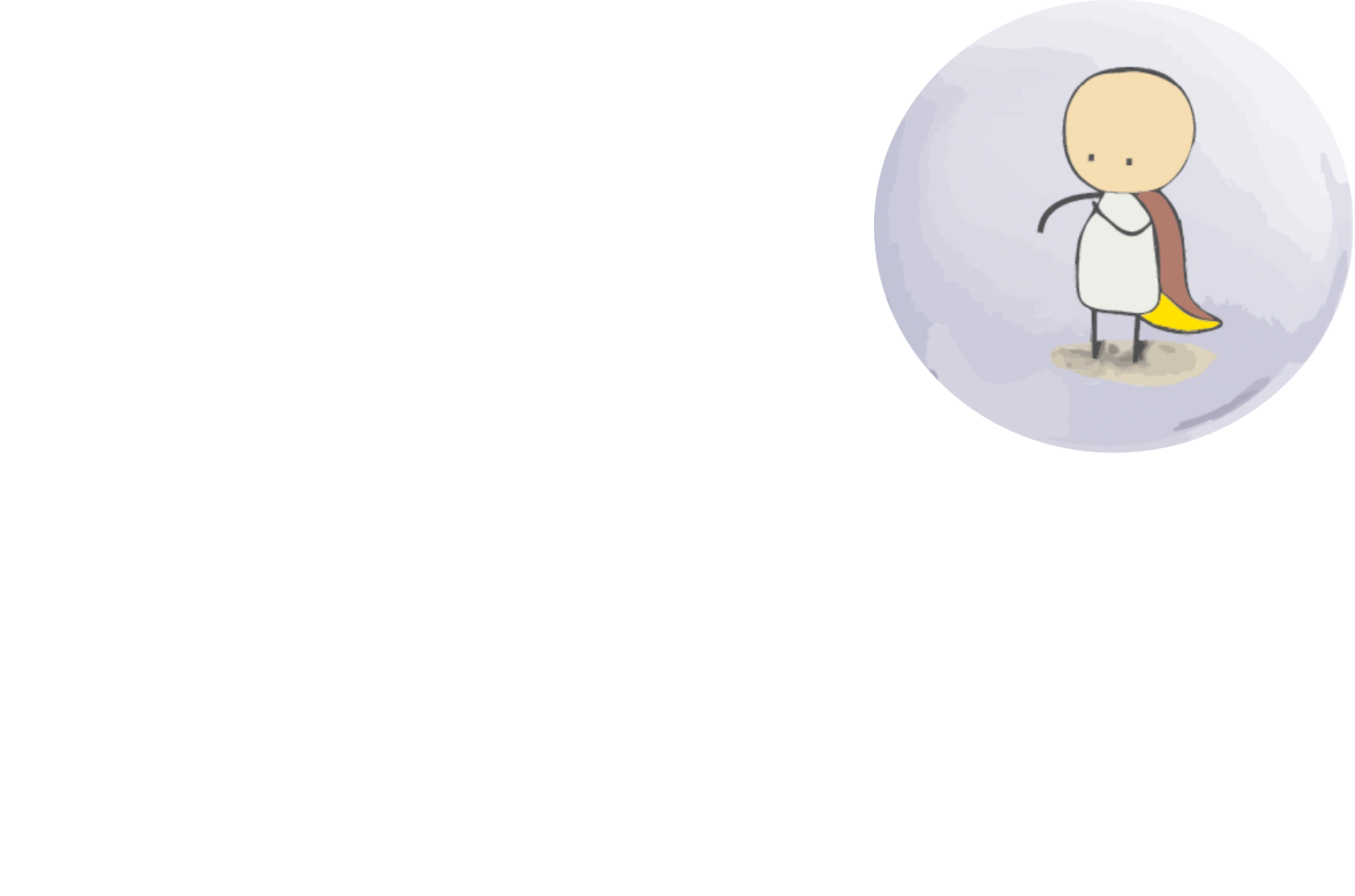}}%
    \put(0.445,0.21){\color[rgb]{0,0,0}\makebox(0,0)[lt]{\lineheight{1.25}\smash{\begin{tabular}[t]{l}\large$V$\end{tabular}}}}%
    \put(0.605,0.17){\color[rgb]{0,0,0}\makebox(0,0)[lt]{\lineheight{1.25}\smash{\begin{tabular}[t]{l}\large$U$\end{tabular}}}}%
    \put(0,0){\includegraphics[width=\unitlength,page=2]{drawing_poster.pdf}}%
    \put(0.48,0.48){\color[rgb]{0,0,0}\makebox(0,0)[lt]{\lineheight{1.25}\smash{\begin{tabular}[t]{l}\large$W$\end{tabular}}}}%
    \put(0,0){\includegraphics[width=\unitlength,page=3]{drawing_poster.pdf}}%
    \put(0.25,0.211){\color[rgb]{0,0,0}\makebox(0,0)[lt]{\lineheight{1.25}\smash{\begin{tabular}[t]{l}\large$\ket{\psi_0}$\end{tabular}}}}%
    \put(0.26,0.122){\color[rgb]{0,0,0}\makebox(0,0)[lt]{\lineheight{1.25}\smash{\begin{tabular}[t]{l}\large$\ket{f}$\end{tabular}}}}%
  \end{picture}%
\endgroup%

	\caption{\label{fig:Alice_Bob}The analogy with quantum communication: first, Alice initializes the quantum computer to be in the state $\ket{\psi_0}$ and chooses an oracle $f$. Then she performs the first unitary gate $V$ on the computer, followed by the controlled gate $U$. As a result, the state of the computer is $ U_f V \ket{\psi_0} $, which she subsequently sends to Bob. Bob is then tasked with choosing a gate $W$ to perform on the state, before measuring it in the computational basis, such that he would optimize the information he obtains on Alice's chosen oracle $f$ or its class $j$.}
\end{figure}

Note that there is another---arguably more illuminating---way of delineating between the ``Alice'' and ``Bob'' roles in the algorithm. One could say that Alice's preparation ends immediately after the oracle query; then Bob obtains an intermediate state, where $ T_f $ in \eqref{rho_Yj_def} is replaced by $ U_f V $ (this intermediate state is denoted $ \rho_Y^2 $ in the next section). Then Bob must choose the measurement basis that would allow him to extract as much information as possible about $J$; effectively, he applies a corresponding unitary $W$ and then measures in the computational basis (see \ref{fig:Alice_Bob}). This point of view may help motivate the next subsection.

\subsection{\label{subsec:discord}Quantum discord and optimal quantum algorithms}
It turns out that a simple construction not only yields another way to derive Holevo's inequality, but also sheds new light on the quantum informational structure of our setting. The idea is as follows: we wish to describe the oracle as a quantum subsystem, in addition to the quantum computer. Moreover, we want the oracle subsystem to be constructed in a manner that would allow us later to ``forget'' about the specific oracle identities $ f \in \mathcal{F} $, and retain only information regarding the oracle classes $ A_j $. Note that this construction---as well as the results obtained hereafter---holds a striking resemblance to a construction performed in the context of Quantum Darwinism~\cite{zwolak2013complementarity}.
Let $ \mathcal{H} = \mathcal{H}_J \otimes \mathcal{H}_F \otimes \mathcal{H}_Y $ be the full Hilbert space, where:
\begin{equation}
	\dim \mathcal{H}_J = \abs{\mathcal{J}} , \quad \dim \mathcal{H}_F = \abs{\mathcal{F}} , \quad \dim \mathcal{H}_Y = N = 2^n.
\end{equation}
Let $ \left\{ \ket{j} \right\}_{j \in \mathcal{J} } $ be an orthonormal basis for $ \mathcal{H}_J $ indexed by the set of oracle classes, and let $ \left\{ \ket{f} \right\}_{f \in \mathcal{F}} $ be an orthonormal basis for $ \mathcal{H}_F $. This notation for the Hilbert spaces, as well as the corresponding notation for states on those Hilbert spaces (see below), follows the convention we have established earlier: a Hilbert space with subscript $X$ has a distinguished orthonormal basis corresponding to the domain of the random variable $X$.
Note this is a \textit{formal} construction, where the states $\ket{j}, \ket{f}$ are not implied to represent an actual physical system. Now, define the classical-classical-classical state $\rho_{JFY}^0$:
\begin{equation}
	\rho_{JFY}^0 \defeq \sum_{j \in \mathcal{J}} \sum_{f \in A_j} p_f \ket{j} \bra{j} \otimes \ket{f} \bra{f} \otimes \ket{ \psi_0 } \bra{ \psi_0 } .
\end{equation}
We say that a state is ``classical-classical-classical'' if it can be written as a mixture of orthonormal basis elements on all three subsystems (generalizing the more common notion of classical-classical states).
In our case, $\rho_{JFY}^0$ consists of three subsystems which represent the oracle class $J$, the oracle identity $F$ and the quantum computer $Y$ in its initial state, $ \ket{\psi_0} $. This state evolves with every step of the quantum algorithm. After applying the unitary $V$ we still have a classical-classical-classical state:
\begin{equation}
	\rho_{JFY}^1 = \sum_{j \in \mathcal{J}} \sum_{f \in A_j} p_f \ket{j} \bra{j} \otimes \ket{f} \bra{f} \otimes V \ket{ \psi_0 } \bra{ \psi_0 } V^\dagger.
\end{equation}
However, the oracle query $U_f$ generates correlations between the ``oracle subsystem'' and the quantum computer:
\begin{equation}
	\rho_{JFY}^2 = \sum_{j \in \mathcal{J}} \sum_{f \in A_j} p_f \ket{j} \bra{j} \otimes \ket{f} \bra{f} \otimes U_f V \ket{ \psi_0 } \bra{ \psi_0 } V^\dagger U_f^\dagger .
\end{equation}
Note that the above state is not necessarily classical-classical-classical, since the states $ \left\{ U_f V \ket{ \psi_0 } \right\}_{f \in \mathcal{F}} $ may not be orthogonal.
Finally, after application of the second unitary $W$ we obtain the state of the system right before the measurement:
\begin{equation}
	\rho_{JFY}^\mathrm{fin} = \sum_{j \in \mathcal{J}} \sum_{f \in A_j} p_f \ket{j} \bra{j} \otimes \ket{f} \bra{f} \otimes T_f \ket{ \psi_0 } \bra{ \psi_0 } T_f^\dagger .
\end{equation}
Since we do not care about the oracle identity $f$, we may trace out the subsystem $F$, to obtain:
\begin{multline}\label{final_state}
	\rho_{JY} = \sum_{j \in \mathcal{J}} \sum_{f \in A_j} p_f \ket{j} \bra{j} \otimes T_f \ket{ \psi_0 } \bra{ \psi_0 } T_f^\dagger = \\
	= \sum_{j \in \mathcal{J}} p_j \ket{j} \bra{j} \otimes \sigma_j .
\end{multline}
Note that $\rho_Y$ defined before is now simply given by taking the partial trace of $ \rho_{JY} $ with respect to the subsystem $J$. At this point we may also trace out the non-measured qubits.

Now, let us compute the \textit{non-optimized quantum discord} of the state $ \rho_{JY} $ with respect to measurement of subsystem $Y$ in the computational basis. This is defined as follows:
\begin{equation}\label{discord_exp}
	D_Y \left( \rho_{JY} ; Z^{\otimes n} \right) = S \left( \rho_Y \right) -S \left( \rho_{JY} \right) + S \left( \rho_J \vert Z^{\otimes n} \right) ,
\end{equation}
where $ \rho_J = \sum_{j \in \mathcal{J}} p_j \ket{j} \bra{j} $ remains unchanged throughout the run of the algorithm. We have the following result (see \Cref{app:discord} for the proof):
\begin{align}\label{Pawel_identity}
	D_Y \left( \rho_{JY} ; Z^{\otimes n} \right) = \chi - I \left( J; Y \right) .
\end{align}
which is nonnegative, as quantum discord always is. In fact, we have found here that the non-negativity of $ D_Y \left( \rho_{JY} ; Z^{\otimes n} \right) $ is equivalent to Holevo's bound. Note that this result was recently derived independently in the context of Bayesian metrology~\cite{Lecamwasam2024relative}, although the non-optimized discord appears there under the name \textit{ensemble relative entropy of coherence}.

However, the more commonly used definition of quantum discord does not fix the measurement basis, but rather it optimizes over all possible measurements:
\begin{equation}
	D_Y \left( \rho_{JY} \right) \defeq \min_\Pi D_Y \left( \rho_{JY} ; \Pi \right) ,
\end{equation}
where $\Pi = \left\{ \Pi_k \right\}_k$ is a projective measurement. A quantum algorithm can be optimal only if it applies unitary gates such that the computational basis is the optimal one, i.e., it yields the minimal oracle-outcome discord (and thus the mutual information obtained is as high as possible). Here one should note that an $f$-independent unitary gate cannot change $ \chi $, but it can increase (or decrease) $ I(J;Y) $. The source of information (about the oracle) in the quantum computer, $ \chi $, as one should expect, stems exclusively from oracle queries.

Let us reiterate the crucial point here. In what follows, $S \left( \rho_Y^i \right)$ denotes the von Neumann entropy of the subsystem $\rho_Y^i = \Tr_{J,F,N} \left( \rho_{JFY}^i \right) $, where $\rho_{JFY}^i$ is the entire system right after the $i$-th step of the algorithm and $N$ denotes the non-measured qubits; $I \left(J; Y_i \right)$ should be understood as the mutual information we would have obtained if we had measured in the computational basis right after the $i$-th step of the algorithm; and $ \ket{\psi_1} \defeq V \ket{\psi_0} $. By changing our chosen $V$ we can set $ \ket{\psi_1} $ to be any normalized pure state. Then, for such $ \ket{\psi_1} $, we get the following state after the third step:
\begin{equation}
	\rho_{JY}^2 \left( \ket{\psi_1} \right) = \sum_{j \in \mathcal{J}} p_j \ket{j} \bra{j} \otimes \left( \sum_{f \in A_j} \frac{p_f}{p_j} U_f \ket{ \psi_1 } \bra{ \psi_1 } U_f^\dagger \right) ,
\end{equation}
where the expression in the parentheses is denoted $ \sigma_j^2 $. Hence, the oracle acts as a map that produces an ensemble $ \left\{ \sigma_j^2 \right\} $ out of the pure state $ \ket{\psi_1} $. In the fourth step, the best we can do is choose a unitary $W$ such that the measurement in the computational basis would be the one with minimal discord. Thus, for any fixed $ \ket{\psi_1} $, we have:
\begin{equation}\label{I_max}
	I_{\max} \left( \ket{\psi_1} \right) = \chi \left( \left\{ p_j, \sigma_j^2 \right\}_{j \in \mathcal{J}} \right) -D_Y \left( \rho_{JY}^2 \right) ,
\end{equation}
where $ I_{\max} \left( \ket{\psi_1} \right) $ denotes the maximally obtainable $ I \left( J; Y \right) $ if we have the state $ \ket{\psi_1} $ after the second step. Note that here $ D_Y \left( \rho_{JY}^2 \right) $ is the \textit{minimal} discord.

Our analysis can be summarized in the following theorem:
\begin{theorem}\label{theorem:optimal}
	For any oracle problem of the form defined in \Cref{oracle_problems} and any fixed value of $ n \geq m $, a single-query quantum algorithm of the form defined in \Cref{algo} that uses $n$ qubits obtains the maximal value of $ I \left( J;Y \right) $ out of all such algorithms (i.e. is \textit{optimal}) iff the following conditions are both satisfied:
	\begin{itemize}
		\item the pre-query unitary $V$ has the property
		\begin{equation*}
			I_{\max} \left( V \ket{\psi_0} \right) = \max_{\ket{\psi_1}} I_{\max} \left( \ket{\psi_1} \right) ,
		\end{equation*}
		where the maximization is over all $n$-qubit states $ \ket{\psi_1} $ ;
		\item the post-query unitary $W$ has the property that $W^\dagger$ maps the computational basis to a basis which yields the minimal discord.
	\end{itemize}
\end{theorem}
Note there is a single-query quantum algorithm that succeeds with probability $1$ iff there exists a state $ \ket{\psi_1} $ such that $ I_{\max} \left( \ket{\psi_1} \right) = H \left( J \right) $. Since $ \chi \left( \rho_{JY}^2 \right) \leq H \left( J \right) $, this state must satisfy the following two conditions simultaneously:
\begin{enumerate}
	\item $ \chi \left( \rho_{JY}^2 \right) $ obtains its maximal bound $ H \left( J \right) $. This is an interesting condition, as there exist finer bounds that use properties of $ \sigma_j^2 $s---see \cite{audenaert2014quantum,shirokov2019upper}. In particular, this occurs iff the states $ \sigma_j^2 $ have orthogonal support~\cite{nielsen2002quantum}. Since $ \left\{ \sigma_j^2 \right\} $ are positive semi-definite, having orthogonal support is equivalent to orthogonality: $ \forall j \neq j' \in \mathcal{J} ,\; \Tr \left( \sigma_j^2 \sigma_{j'}^2 \right) = 0 $.
	
	\item $ D_Y \left( \rho_{JY}^2 \right) = 0 $, i.e. the state $ \rho_{JY}^2 $ is not only classical-quantum, but also quantum-classical. This implies the state is classical-classical, and this scenario occurs iff the states $ \sigma_j^2 $ pairwise commute.
\end{enumerate}
For any two positive semi-definite matrices $ \rho $ and $\sigma$, we have that $ \Tr \left( \rho \sigma \right) = 0 $ implies $ \rho \sigma = \sigma \rho = 0 $, and in particular $ \left[ \rho, \sigma \right] =0 $. Thus, the first condition implies the second one (but not the other way around).

Indeed, if the first condition holds, then a simple way to construct $W$ would be to map the orthogonal support subspaces to the computational basis. Further details and proof of optimality appear in \Cref{app:prob_1}. Moreover, even if the first condition does not hold but the second one does, the optimal $W$ (for fixed $V$) is one that simultaneously diagonalizes all of the states $ \sigma_j^2 $. This is evident from the observations above since $ I \left( J;Y \right) $ obtains its upper bound $\chi$, which cannot be changed by $W$. This fact remains true even if we are allowed to use arbitrary POVMs instead of projective measurements, since Holevo's bound is true for POVMs as well.
\begin{remark}
	Suppose $V$ is a monomial matrix, i.e. having exactly one nonzero entry in each row and column. Such a matrix cannot generate superpositions: it maps computational basis states to computational basis states multiplied by some phase. Usually $U_f$ has the stronger property of being a permutation matrix; this assumption is necessary if we think of computational basis states as being classical. In this scenario the matrices $ \sigma_j^2 $ are all diagonal in the computational basis, hence commuting. By the above discussion, $ W = I$ is optimal; and since the phase is insignificant, such an algorithm can be executed by a classical computer. These facts can be interpreted as saying that if the oracle query is performed classically, then quantum post-processing cannot be used to increase the amount of available classical information. Another consequence is that a quantum algorithm with monomial $V$ can never outperform the optimal classical algorithm. Indeed, in every quantum algorithm we take as an example in \Cref{sec:analysis}, $V$ is a generalized complex Hadamard matrix, i.e. a unitary matrix whose entries all have the same absolute value. Such a matrix is as far from being monomial as possible.
\end{remark}

To conclude this subsection, we note that the above theorem neatly characterizes optimal single-query quantum algorithms, and via the correspondence described in \ref{non-adaptive} it also applies to multiple-query non-adaptive algorithms with a fixed number of queries. However, it may be challenging to find the optimal $ \ket{\psi_1} $ in practice for a general oracle problem (even for single-query algorithms).

Because of this practical difficulty, the theorem as stated is largely conceptual. It reveals that extracting maximal information is mathematically equivalent to minimizing quantum discord, thus translating an abstract computational condition into the language of fundamental physical resources. While this conceptual lens does not directly yield novel, standalone single-query algorithms, the underlying mathematical framework carries significant practical utility. As we discuss in \Cref{subsec:QLE}, our mutual-information perspective provides the exact objective function needed to optimize iterative subroutines in hybrid quantum--classical algorithms, where accumulating partial information over multiple steps is more relevant than achieving a high single-shot success probability.

\subsection{Coherence and a lower bound on the mutual information}
Recall that $ I \left( J ; Y \right) = H \left( Y \right) - H \left( Y \vert J \right) $. A simple observation (see \Cref{app:lower_bound}) shows:
\begin{equation}
	H \left( Y \right) \geq S \left( \rho_Y \right)
\end{equation}
where the LHS is the Shannon entropy of the bit string obtained by measuring the first $k$ qubits, and the RHS is the von Neumann entropy of the reduced state obtained by tracing out the $n-k$ non-measured qubits.
Moreover, the difference between the former and the latter is known to equal the \textit{relative entropy of coherence}~\cite{baumgratz2014quantifying,xi2015quantum}: 
\begin{equation}\label{coherence}
	C \left( \rho_Y \right) = H \left( Y \right) -S \left( \rho_Y \right) \geq 0 .
\end{equation}
Thus, just as the non-optimized discord measures the ``failure'' of $ I \left( J; Y \right) $ to obtain its upper bound, so does the relative entropy of coherence measure its failure to obtain the lower bound. Writing the lower and upper bounds together, we have:
\begin{equation}\label{mutual_information_bounds}
	S \left( \rho_Y \right) - H \left( Y \vert J \right) \leq I \left( J ; Y \right) \leq S \left( \rho_Y \right) - \sum_{j \in \mathcal{J}} p_j S \left( \sigma_j \right) .
\end{equation}
The sum $ \Im_{Z^{\otimes n}} \left( \rho_{JY} \right) \defeq C \left( \rho_Y \right) + D_Y \left( \rho_{JY} ; Z^{\otimes n} \right) $ is known as the \textit{irrealism}~\cite{Bilobran_2015,Dieguez_2018_Information_reality,costa2020information,Paiva2023Coherence} of measuring the $Y$ subsystem in the computational basis. Note that $ \Im_{Z^{\otimes n}} \left( \rho_{JY} \right) $ vanishes iff the lower and upper bounds in \eqref{mutual_information_bounds} are both saturated.

\subsection{\label{subsec:QLE}Application to hybrid quantum--classical schemes}
While our results are not directly applicable to adaptive quantum algorithms, they can inform the optimization of hybrid quantum--classical schemes. Consider the task of learning an unknown Hamiltonian $H_{\text{true}}$ from a finite set of candidates ${H_1, H_2, \dots, H_N}$. The learner can access $H_{\text{true}}$ only through its unitary evolution, $U(t) = e^{-i t H_{\text{true}} / \hbar}$. A hybrid quantum--classical algorithm alternates between quantum and classical stages. Each quantum stage consists of a circuit involving a single application of $U(t_i)$ for some evolution time $t_i$, which may vary between iterations. Between successive quantum stages, classical processing uses the measurement outcomes to update the stored information about $H_{\text{true}}$ and to determine the parameters of the next quantum stage. After a finite number of iterations, the algorithm outputs an estimate of $H_{\text{true}}$.

Quantum Likelihood Estimation (QLE) provides a concrete example of such a hybrid scheme~\cite{granade2012robust,wiebe2014hamiltonian,Wiebe2014Certification,nphys4074,qi2019determining,dutt2021active}. In its standard form, QLE maintains a probability distribution representing the information accumulated so far about $H_{\text{true}}$. However, only the evolution time $t_i$ is adaptively chosen between iterations, while the pre- and post-evolution unitaries are kept fixed throughout. In a recent paper~\cite{levi2025optimalquantumlikelihoodestimation} made possible by the current work, some of us proposed that each quantum iteration can be modeled as a single-query quantum algorithm, and the results developed here were applied to optimize the information gain about $H_{\text{true}}$ at each step. The resulting optimized version of QLE achieves substantially faster convergence, requiring far fewer iterations to learn $H_{\text{true}}$ than the original algorithm. This shows that the proposed framework could have some practical ramifications in addition to its conceptual contributions.

\section{\label{sec:analysis}Information-theoretic analysis of standard quantum algorithms}
In this section we demonstrate our approach by studying several examples of quantum algorithms. The information-theoretic figures we have discussed are computed and presented in detail, using a combination of analytic and numerical methods. Specifically, while we provide exact analytical derivations wherever possible, we utilize numerical simulations for cases where closed-form expressions for the subsystem entropies are mathematically intractable. These numerics are intended to empirically illustrate the dynamics of discord and coherence predicted by our framework, demonstrating that physical resources can be tracked even when exact analytical formulas evade us.

\subsection{\label{subsection:DJ}Simulation results for Deutsch--Jozsa}
Starting with the Deutsch--Jozsa algorithm, we have run simulations for $ k \leq 4 $, where $k$ denotes the number of input bits for the oracle functions (which is also the number of measured qubits). In particular, the simulation (a MATLAB script) goes over all possible oracle functions and emulates the algorithm's run for each, computing the state of the quantum computer (pre-query, post-query and final) and the probabilities for every possible measurement outcome. The states are mixed together to find $\rho_Y^i$, the probabilities are used to find the probability distribution of $Y$ conditioned on $J$, and these are then used to calculate the required information-theoretic quantities. The results are displayed in \ref{DJ_Simulation}. Note we have taken the ``partition-uniform'' probability distribution over the oracles.

\begin{table}
 	\caption{\label{DJ_Simulation}Information-theoretic quantities for the Deutsch--Jozsa algorithm, found by simulating the algorithm's run over all possible inputs for $k \leq 4$. The three parts of the table depict quantities obtained by measuring the output qubits right before the oracle query (first row), immediately after the query (second row), or at the end of the algorithm (third row). Here $D_Y^i \defeq D_Y \left( \rho_{JY}^{i} ; Z^{\otimes n} \right)$ for $i=1,2$, and $D_Y \defeq D_Y \left( \rho_{JY} ; Z^{\otimes n} \right)$.}
 \begin{ruledtabular}
 \begin{tabular}{ c c c c c c c c }
	$ k $ & $ H \left( Y_1 \right) $ & $ S \left( \rho_Y^1 \right) $ & $ C \left( \rho_Y^1 \right) $ & $ H \left( Y_1 \vert J \right) $ & $ \chi_1 $ & $ I \left( J ; Y_1 \right) $ & $ D_Y^1 $\\
	\midrule
	1 & 1 & 0 &	1 & 1 & 0 & 0 & 0 \\ 
	2 & 2 & 0 & 2 & 2 & 0 & 0 & 0 \\ 
	3 & 3 & 0 & 3 & 3 & 0 & 0 & 0 \\
	4 & 4 & 0 & 4 & 4 & 0 & 0 & 0 \\
	\midrule
	$ k $ & $ H \left( Y_2 \right) $ & $ S \left( \rho_Y^2 \right) $ & $ C \left( \rho_Y^2 \right) $ & $ H \left( Y_2 \vert J \right) $ & $ \chi_2 $ & $ I \left( J ; Y_2 \right) $ & $ D_Y^2 $\\
	\midrule
	1 & 1 & 1		&	0 &  1 & 1 & 0 & 1 \\ 
	2 & 2 & 1.7925 & 0.2075 & 2 & 1 & 0 & 1 \\ 
	3 & 3 & 2.4037 & 0.5963 & 3 & 1 & 0 & 1 \\
	4 & 4 & 2.9534 & 1.0466 & 4 & 1 & 0 & 1 \\
	\midrule
	$ k $ & $ H \left( Y \right) $ & $ S \left( \rho_Y \right) $ & $ C \left( \rho_Y \right) $ &$ H \left( Y \vert J \right) $ & $ \chi $ & $ I \left( J ; Y \right) $ & $ D_Y $ \\ 
	\midrule
	1 & 1 		& 1 	& 0 & 0 & 1 & 1 & 0 \\ 
	2 & 1.7925 & 1.7925 & 0 & 0.7925 & 1 & 1 & 0 \\ 
	3 & 2.4037 & 2.4037 & 0 & 1.4037 & 1 & 1 & 0 \\
	4 & 2.9534 & 2.9534 & 0 & 1.9534 & 1 & 1 & 0 \\
 \end{tabular}
 \end{ruledtabular}
\end{table}

As seen above, the simulations demonstrate that the discord $ D_Y \left( \rho_{JY} ; Z^{\otimes n} \right) $ vanishes for Deutsch--Jozsa. For example, with $k = 4$ we obtain $ S \left( \rho_Y \right) \approx 2.95 $, while for the balanced and constant functions respectively we have $ S \left( \rho_Y^b \right) \approx 3.91 $ and $ S \left( \rho_Y^c \right) = 0 $. Since the Deutsch--Jozsa algorithm always succeeds, we have $ I \left( J; Y \right) = 1 $. Hence, the discord is $ 2.95 -\frac{1}{2}\cdot 3.91 -\frac{1}{2}\cdot 0 -1 = 0 $.

In fact, a closer look into the DJ algorithm reveals the following interesting structure. Upon initialization, the von Neumann entropy and mutual information are both zero; and this is still the case after applying $V$. After the oracle query, the von Neumann entropies of $ \rho_Y, \sigma_j $ already obtain their final values, i.e. $S \left( \rho_Y^2 \right) = S \left( \rho_Y^\mathrm{fin} \right)$ and $ S \left( \sigma_j^2 \right) = S \left( \sigma_j^\mathrm{fin} \right) $. Thus, $\chi$ obtains its final value, which equals $ H \left( J \right) $. $I \left( J; Y_2 \right)$ obtains some value $\epsilon$.
After application of the second unitary $W$ and before the measurement, the mutual information $ I \left( J; Y \right) $ obtains its maximal possible value, i.e. $ I \left( J; Y \right) = \chi = H \left( J \right) $, such that the discord takes its minimal value, which is zero.

\subsection{\label{subsection:BV}Information-theoretic quantities for Bernstein--Vazirani}
While the Bernstein--Vazirani algorithm itself is essentially identical to Deutsch--Jozsa, the promise on the function $f$ is different, as well as the partition $ \mathcal{J} $. In particular, we are promised that $f$ is of the form $ f_a \left( x \right) \defeq a \cdot x $ where $ a \in \left\{ 0,1 \right\}^n $. The objective is to determine the parameter $a$ of $f_{a}$ exactly, i.e. the classes are singletons; hence $ J = F $. These properties make this algorithm particularly well-behaved, allowing us to compute the information-theoretic quantities analytically. The findings are summarized in \ref{BV}.
\begin{table}
	\caption{\label{BV}Information-theoretic quantities for the Bernstein--Vazirani algorithm, computed analytically for any value of $n$.}
	\begin{ruledtabular}
		\begin{tabular}{ c c c c }
			Quantity & Pre-query & Post-query & Final \\
			\midrule
			$H\left(Y\right)$ & $n$ & $n$ & $n$ \\
			$S\left(\rho_Y \right)$ & $0$ & $n$ & $n$ \\
			$C \left( \rho_Y \right)$ & $n$ & $n$ & $0$ \\
			$H\left( Y \vert F \right)$ & $n$ & $n$ & $0$ \\
			$I\left( F; Y \right)$ & $0$ & $0$ & $n$ \\
			$ D_Y \left( \rho_{FY} ; Z^{\otimes n} \right) $ & $0$ & $n$ & $0$ \\
		\end{tabular}
	\end{ruledtabular}
\end{table}
Note that since $ \sigma_j $ are all pure states, we have $ \chi = S \left( \rho_Y \right) $. As in the Deutsch--Jozsa algorithm, one can observe that the coherence $ C \left( \rho_Y \right) $ and conditional entropy $ H \left( Y \vert J \right) $ are initially large but are decreased by the second unitary $W$. In contrast, the von Neumann entropy and discord have initial values zero but are increased to $n$ by the oracle query. The second unitary $W$ can be said to ``trade off'' the discord in exchange for mutual information, which obtains the maximal value $n$ in the final state.
The detailed computation of these quantities can be found in \Cref{app:BV}.

\subsection{\label{subsection:HSP}The Shor--Kitaev algorithm}
Suppose we are given a group $G$, a set $X$ and an unknown function $f : G \rightarrow X$ that \textit{hides} a subgroup $H \subset G$. That is, we have the following commutative triangle:
\begin{equation}
	\begin{tikzcd}
		G \arrow[rr,"f"] \arrow[ddr,two heads,"\pi"] && X \\ && \\
		& G/H \arrow[uur,hook]
	\end{tikzcd}	
\end{equation}
In this case, $f$ is the composition of the canonical projection $ \pi: G \rightarrow G/H $ with any one-to-one map (of sets) $ G/H \hookrightarrow X $ and we say that $f$ hides the subgroup $H$. We are given oracle access to the function $f$, and the task is to find the subgroup $H$ using as few queries as possible. This framework, known as the hidden subgroup problem (HSP), encompasses several well-known problems including Simon's problem, the discrete logarithm, period and order finding. If $G$ is finite and abelian, the Shor--Kitaev algorithm solves the HSP efficiently using a quantum Fourier transform~\cite{nielsen2002quantum}. In fact, in the case of order finding, this algorithm forms the quantum subroutine in Shor's prime factorization algorithm. Since this generic algorithm is non-adaptive, our framework applies. In fact, it is not even a proper multiple-query algorithm; rather it is the same single-query algorithm executed several times to learn the probability distribution of $Y$ with greater accuracy.

We now describe the single-query routine. Let $ \mathcal{F} $ be the set of allowed functions $f$, and $ \mathcal{J} $ is the partition of $ \mathcal{F} $ into classes such that functions that hide the same subgroup $H$ are in the same class. Thus the indices $ j \in \mathcal{J} $ correspond to subgroups $ H_j \subset G $.
Assuming the standard ``bit oracle'' for the function $f$, the quantum computer should consist of two registers that correspond to elements of $G$ and $X$ respectively. The joint state of the oracle and computer at each step $i$ is of the form:
\begin{equation}
	\rho_{JFY}^i = \sum_{j \in \mathcal{J}} \sum_{f \in A_j} p_f \ket{j} \bra{j} \otimes \ket{f} \bra{f} \otimes \ket{ \psi_i \left( f \right) } \bra{\psi_i \left( f \right)} 
\end{equation}
where $ A_j $ is the set of all functions that hide the subgroup $ H_j $ of $G$ and $ \ket{ \psi_i \left( f \right) } $ is the state of the computer at step $i$. In particular, right before the query we have:
\begin{equation}
	\ket{\psi_1} = \frac{1}{\abs{G}^{1/2}} \sum_{g \in G} \ket{g} \ket{0} 
\end{equation}
After the query:
\begin{equation}
	\ket{\psi_2 \left( f \right)} = \frac{1}{\abs{G}^{1/2}} \sum_{g \in G} \ket{g} \ket{f \left( g \right)}
\end{equation}
Choosing a representative $ s \in gH_j $ for each coset $ gH_j \in G/H_j $, we can write:
\begin{equation}
	\ket{\psi_2 \left( f \right)} = \frac{1}{\abs{G}^{1/2}} \sum_{s \in G/H_j} \left( \sum_{h \in H_j} \ket{s+h} \right) \otimes \ket{f \left( s \right)}
\end{equation}
using the fact that $ f \left( g \right) $ is constant on cosets. The usual description of the algorithm states that here the second register should be measured and discarded (and the measurement outcome is not used). By the principle of deferred measurement, we can move this measurement to the end of the algorithm and obtain the exact same action of the circuit.

Finally, we perform a quantum Fourier transform on the first register. This has the following effect:
\begin{equation}
	\frac{1}{\abs{H_j}^{1/2}} \sum_{h \in H_j} \ket{s+h} \mapsto \sqrt{\frac{\abs{H_j}}{\abs{G}}} \sum_{\chi_g \in H_j^\perp} \chi_g \left( s \right) \ket{g}
\end{equation}
where $ \chi_g \in \hat{G} = \mathrm{Hom} \left( G, \mathrm{U} \left( 1 \right) \right) $ is a character of $G = \Pi_{i=1}^M \left( \mathbb{Z} / p_i  \mathbb{Z} \right)$ defined by
\begin{equation}
	\chi_g \left( l_1, \ldots l_M \right) \defeq \Pi_i e^{2\pi i l_i g_i/p_i} \quad \mathrm{for} \quad g = \left( g_1, \ldots, g_M \right) ;
\end{equation}
and $ H_j^\perp $ is the subgroup of $\hat{G}$ comprising the characters $\chi$ such that $ \mathrm{Res}_{H_j}^G \left( \chi \right) = \chi_e $, i.e. the characters that restrict to the trivial character on $H_j$.
So we have:
\begin{equation}
	\ket{\psi_{\mathrm{fin}} \left( f \right)} = \frac{\sqrt{\abs{H_j}}}{\abs{G}} \sum_{s \in G/H_j} \left( \sum_{\chi_g \in H_j^\perp} \chi_g \left( s \right) \ket{g} \right) \otimes \ket{f \left( s \right)} .
\end{equation}

Now we wish to analyze the information-theoretic quantities. The random variable $Y$ refers to the outcome of measuring only the first register in the computational basis (i.e. the one corresponding to elements of $G$). The pertinent matrices $ \rho_Y^i $ are:
\begin{equation}
	\rho_Y^i = \sum_{j \in \mathcal{J}} \sum_{f \in A_j} p_f \Tr_{O} \left( \ket{ \psi_i \left( f \right) } \bra{\psi_i \left( f \right)} \right)
\end{equation}
where $ \Tr_{O} $ denotes partial trace over the output register. Let us perform the computation for each of the three steps.

\subsubsection{Pre-query}
After discarding the output register, the state of the computer right before the query is a ``democratic superposition'', hence:
\begin{equation}
	C \left( \rho_Y^1 \right) = H \left( Y_1 \vert J \right) = H \left( Y_1 \right) = \log \abs{G}
\end{equation}
and of course, we have 
\begin{equation}
	S \left( \rho_Y^1 \right) = \chi_1 = I \left( J; Y \right) = D_Y \left( \rho_{JY} ; G \right) = 0 ,
\end{equation}
where $ D_Y \left( \rho_{JY} ; G \right) $ is the discord of $ \rho_{JY} $ with respect to measuring the first register of computer in the computational basis, which here corresponds to the group $G$.

\subsubsection{Post-query}\label{subsub:HSP_post_query}
The query does not change the diagonal entries of $ \sigma_j^2 $, so we have $ H \left( Y_2 \vert J \right) = H \left( Y_2 \right) = \log \abs{G} $ and $ I \left( J; Y_2 \right) = 0$.
It turns out that for all $j,j' \in \mathcal{J}$, we have $ \left[ \sigma_j^2, \sigma_{j'}^2 \right] = 0$, while $ \Tr \left( \sigma_j^2 \sigma_{j'}^2 \right) \neq 0 $; i.e., out of the two conditions for optimality described in \ref{subsec:discord}, the second condition is satisfied while the first one is not. This means that the minimized discord $ D_Y \left( \rho_{JY}^2 \right) $ vanishes. It also means that the matrices $ \sigma_j^2 $ and $\rho_Y^{2}$ are all mutually diagonalizable; in fact, their common eigenbasis is $ \left\{ \ket{v_\chi} \right\}_{\chi \in \hat{G}} $, which are the column vectors of the inverse quantum Fourier transform over $G$ (equivalently---the complex conjugates of the rows of the character table of $G$).
This is also the basis in which the minimal discord (i.e. zero) is obtained, and indeed the final unitary $W$ is the quantum Fourier transform, as we would have expected from a ``good'' algorithm.

These insights allow us to proceed with relative ease and generality. We find that $ S \left( \sigma_j^2 \right) = \log \abs{G} -\log \abs{H_j} $. All other quantities depend on finding $ S \left( \rho_{Y}^{2} \right) $: we have $ C \left( \rho_{Y}^{2} \right) = \log \abs{G} - S \left( \rho_{Y}^{2} \right) $ and $ D_Y \left( \rho_{JY}^2 ; G \right) = \chi_2 \left( \left\{ \sigma_j^2; p_j \right\} \right) = S \left( \rho_{Y}^{2} \right) -\log \abs{G} +\sum_{j \in \mathcal{J}} p_j \log \abs{H_j} $. Unfortunately, there seems to be no concise expression for $ S \left( \rho_{Y}^{2} \right) $; the best we can do here is provide expressions for the eigenvalues of $ \rho_{Y}^{2} $:
\begin{equation}\label{eigenvalues}
	\lambda_\chi = \frac{1}{ \abs{G}} \sum_{j \in \mathcal{J} \, : \, \chi \in H_j^\perp} p_j \abs{H_j} .
\end{equation}
If $J$ is uniformly random, we can also write them down as:
\begin{equation}
	\lambda_\chi = \frac{1}{\sqrt{\abs{G}} \abs{\mathcal{J}}} \braket{ c \vert v_\chi } ,
\end{equation}
where $ \ket{ c } = \sum_{g \in G} c_g \ket{g} $ and $c_g$ is defined to be the number of subgroups $H_j$ of which $g$ is an element. This way, the eigenvalues can be computed directly from the character table by multiplying it from the left with a row vector. For proofs and further details see \Cref{app:SK}.

\subsubsection{\label{subsub:HSP_final}Final state (pre-measurement)}
As mentioned before, after the quantum Fourier transform the matrices $ \sigma_j $ are all diagonal, hence so is $ \rho_Y $. Thus we have $ C \left( \rho_Y \right) = 0 $ and $ H \left( Y \right) = S \left( \rho_Y \right) = S \left( \rho_Y^2 \right) $, and the latter can be computed using the eigenvalues \eqref{eigenvalues}. We may easily compute:
\begin{equation}
	H \left( Y \vert J \right) = \sum_{j \in \mathcal{J}} p_j S \left( \sigma_j \right) = \log \abs{G} - \sum_{j \in \mathcal{J}} p_j \log \abs{H_j} .
\end{equation}
Moreover, $	S \left( \sigma_j \right) = S \left( \sigma_j^2 \right) = \log \abs{G} -\log \abs{H_j}$, and $
	\chi \left( \left\{ \sigma_j; p_j \right\} \right) = \chi \left( \left\{ \sigma_j^2; p_j \right\} \right) $.
Since $ D_Y \left( \rho_{JY} ; G \right) = D_Y \left( \rho_{JY} \right) = 0 $, we also have $ I \left( J; Y \right) = \chi \left( \left\{ \sigma_j; p_j \right\} \right) $. Note that $ I \left( J; Y \right) $ obtains both its lower and upper bounds. However, $ I \left( J; Y \right) < H \left( J \right) $ since the $ \sigma_j $ do not have orthogonal support.

To conclude this subsection, we discuss the information-theoretic quantities obtained when performing $t$ queries. We only do this for the final stage of the algorithm. We thus consider $\left( \sigma_j \right)^{\otimes t} $ instead of $ \sigma_j $, and define $ \rho_Y^{(t)} \defeq \sum_{j \in \mathcal{J}} p_j \left( \sigma_j \right)^{\otimes t} $. These matrices are still diagonal in the tensor product basis, so $ C \left( \rho_Y^{(t)} \right) =0 $.
Let $Y_t$ denote the outcome of measuring all $t$ input registers at the end of the algorithm. As it turns out, the conditional entropy is simply multiplied by $t$, i.e. $ H \left( Y_t \vert J \right) = t\cdot H \left( Y \vert J \right) $. Describing $ H \left( Y_t \right) = S \left( \rho_Y^{(t)} \right) $ is slightly more complicated than before. The tensor product basis is indexed by tuples of $t$ irreducible representations of $G$, and we have:
\begin{equation}\label{HSP_t_queries_eig}
	\lambda \left( \chi_1, \ldots, \chi_t \right) = \sum_{j \in \mathcal{J} \, : \, \chi_1, \ldots, \chi_t \in H_j^\perp} p_j \left( \frac{\abs{ H_j }}{\abs{G}} \right)^t .
\end{equation}
As in the single-query case, we have vanishing discord, implying
\begin{equation}\label{HSP_t_queries_mutual_info}
	I \left( J; Y_t \right) = S \left( \rho_Y^{(t)} \right) -t\cdot H \left( Y \vert J \right) = \chi \left( \left\{ \left( \sigma_j \right)^{\otimes t}; p_j \right\} \right) .
\end{equation}
While deriving a single, closed-form analytical expression for $ S \left( \rho_Y^{(t)} \right) $ that holds universally across all arbitrary groups $G$ is complicated, the exact values for any specific HSP instance can be computed algorithmically using \eqref{HSP_t_queries_eig} (as we demonstrate for Simon's algorithm in \Cref{subsection:Simon}).
Furthermore, even without explicit computation, one can observe general asymptotic behaviors: we expect $ S \left( \rho_Y^{(t)} \right) $ to grow at least linearly with $t$, since the mutual information $ I \left( J; Y_t \right) $ must be monotonically non-decreasing in $t$. On the other hand, we have $ I \left( J; Y_t \right) \leq H \left( J \right) $, so for large values of $t$ the entropy $ S \left( \rho_Y^{(t)} \right) $ cannot grow faster than a linear function. For any instance of the HSP, analyzing these quantities can be used to find the number of queries needed to obtain any desired amount of mutual information, i.e. $t$ such that $ I \left( J; Y_t \right) \geq H \left( J \right) - \varepsilon $.

It is worth noting that while traditional methods like the Pretty Good Measurement (PGM)~\cite{Barnum2002Reversing} are highly effective for bounding the success probability and estimating asymptotic query requirements for state discrimination, analyzing $S(\rho_Y^{(t)})$ provides a complementary perspective. Specifically, rather than bounding discrimination error, our approach tracks the exact accumulation of partial mutual information per query. This is particularly useful in regimes where query counts are strictly limited and perfect discrimination is impossible.

\subsubsection{\label{order_finding}Order finding}
There is a subtlety in the description of order finding (hence also factorization) within this general framework. Let us briefly describe this problem: for $a,N$ positive integers with $ \mathrm{gcd} \left( a, N \right) = 1 $, we define the order of $a$ modulo $N$ as the smallest positive integer $r$ such that $ a^r = 1 \left( \mathrm{mod} \, N \right) $. The task is to find $r$ for input $a$ and $N$, given an oracle that computes $ g \left( x \right) = ax \left( \mathrm{mod} \, N \right) $. Since $a$ and $N$ are given to us directly as inputs, this is not even a proper oracle problem; indeed, even a classical algorithm can solve it without performing any queries. Nevertheless, this is still an interesting problem to study in terms of time complexity (i.e. the total number of computation steps, rather than counting only oracle queries). As it turns out, the path towards an efficient quantum algorithm for this problem goes through defining a composite oracle out of the basic one that merely computes multiplication. What this composite oracle does is compute \textit{modular exponentiation}:
\begin{equation}
	U_a \ket{x} \ket{y} = \ket{x} \ket{a^x y \left( \mathrm{mod} \, N \right)} ,
\end{equation}
which is an oracle for the function $ f \left( x \right) = a^x \left( \mathrm{mod} \, N \right) $. Since $f$ hides the subgroup $ H = r \mathbb{Z} $ of $ G = \mathbb{Z} $, we get an instance of the hidden subgroup problem, and it can be solved using a quantum Fourier transform. Still, this problem is uninteresting in terms of query complexity; even if we are not given $a$ as an input, a single classical query computes it: $ f \left( 1 \right) = a $. Thus, after this query the mutual information $ I \left( F; Y \right) $ would already obtain its maximal value. Indeed, the advantage of Shor's algorithm does not lie in its ability to extract as much information about the function $f$ using few queries; rather, it is about extracting the piece of information that we care about (the order $r$) without requiring many additional steps of computation. Therefore, one should keep in mind that thinking about the problems of order finding and integer factorization purely in terms of query complexity may not be a useful perspective.

That said, the quantum algorithms for these problems contain a key subroutine which \textit{can} be naturally considered as solving an oracle problem, i.e. phase estimation. We study this important example in \ref{subsection:phase}.

\subsection{\label{subsection:Simon}Information-theoretic quantities in Simon's algorithm}
Recall the description of Simon's problem:
We are given oracle access to a function $f:\left\{ 0,1\right\} ^{n}\rightarrow\left\{ 0,1\right\} ^{n}$, with the promise that for some unknown $s\in\left\{ 0,1\right\} ^{n}$, we have $f\left(x\right)=f\left(y\right)$ iff $x\oplus y\in\left\{ 0^{n},s\right\} $. The goal is to identify $s$. This is an instance of the hidden subgroup problem with $ G = \left( \mathbb{Z} / 2 \mathbb{Z} \right)^n $, where the subgroup is promised to be of the form $ H_s = \left\{ 0, s \right\} $. Note that $s$ may be zero, in which case $ H_s $ has only one element. We consider the uniform distribution and apply the results of the previous section.

The information-theoretic quantities after a single query are summarized in \ref{Simon_NU}. Details of the computation appear in \Cref{app:Simon}. Note that the evolution of the information-theoretic quantities throughout the algorithm somewhat differs from those of the Bernstein--Vazirani algorithm (see \ref{BV}). In Simon's algorithm the intermediate coherence vanishes, whereas in Bernstein--Vazirani it is maximal. Moreover, the amount of information extracted in one query of Simon's algorithm is only of the order of a single bit, while in Bernstein--Vazirani it is $n$. This is reflected directly in comparing the last three rows: indeed, replacing $1$ by $n$ in the values for $ H \left( Y \vert F \right) $, the mutual information and the discord in Simon's algorithm would provide the appropriate values for Bernstein--Vazirani.
\begin{table}
	\caption{\label{Simon_NU}Information-theoretic quantities for a single query of Simon's algorithm, computed analytically for any value of $n$, up to $ O \left( 2^{-n} \right) $.}
	\begin{ruledtabular}
		\begin{tabular}{ c c c c }
			Quantity & Pre-query & Post-query & Final \\
			\midrule
			$H\left(Y\right)$ & 			$n$ & $n$ & $n$ \\
			$S\left(\rho_Y \right)$ & 		$0$ & $n$ & $n$ \\
			$C \left( \rho_Y \right)$ & 	$n$ & $0$ & $0$ \\
			$H\left(Y\vert J\right)$ &		$n$ & $n$ & $n-1$ \\
			$ \chi $				&		$0$ & $1$ & $1$ \\
			$I\left( J; Y \right)$ & 		$0$ & $0$ & $1$ \\
			$D_Y\left(\rho_{JY};G\right)$&	$0$ & $1$ & $0$ \\
		\end{tabular}
	\end{ruledtabular}
\end{table}

These single-query quantities highlight a key advantage of measuring performance via mutual information. After a single query in Simon's algorithm, the probability of successfully identifying the hidden string $s$ is exponentially small (roughly $2^{-n+1}$). A traditional success-probability metric would therefore suggest that virtually no computational progress has been made. However, our framework reveals that $1$ bit of mutual information about the oracle class has been concretely extracted.

Combining analytic and numerical methods, we can also find the information-theoretic quantities for $t$ queries, at the final step of the algorithm, for sufficiently small values of $n$, $t$. The figures thus obtained are depicted in \ref{Simon_t_queries}.
\begin{table}
	\caption{\label{Simon_t_queries}Information-theoretic quantities at the final step of Simon's algorithm, for $2 \leq n \leq 4$ and $ t \leq 24 / n $. Since simulating the run of the algorithm over all possible inputs is intractable, we have utilized the analytical findings of \ref{subsub:HSP_final}. Numerical computations were used to find the eigenvalues of $ \rho_Y^{(t)} $ by evaluating \eqref{HSP_t_queries_eig}, to compute the entropy $ H \left( Y_t \right) = S \left( \rho_Y^{(t)} \right) $ using these eigenvalues, and then use it to find $ I \left( J ; Y_t \right) = \chi $ via \eqref{HSP_t_queries_mutual_info}. Note we have assumed the identities $ C \left( \rho_Y^{(t)} \right) = 0 $ and $ D_Y = 0 $. It is noteworthy that for fixed $n$, $ H \left( Y_t \right) $ grows slightly faster than $ H \left( Y_t \vert J \right) $ (which grows linearly with $t$), thus increasing the mutual information.}
	\scriptsize
	\begin{ruledtabular}
		\begin{tabular}{ c c c c c c c c }
			$ n $ & $t$ & $ H \left( Y_t \right) $ & $ C \left( \rho_Y^{(t)} \right) $ &$ H \left( Y_t \vert J \right) $ & $ \chi $ & $ I \left( J ; Y_t \right) $ & $ D_Y $ \\ 
			\midrule
			2 & 1 & 1.8802 & 0 & 1.25 & 0.6302 & 0.6302 & 0 \\
			2 & 2 & 3.6157 & 0 & 2.5 & 1.1157 & 1.1157 & 0 \\
			2 & 3 & 5.2062 & 0 & 3.75 & 1.4562 & 1.4562 & 0 \\
			2 & 4 & 6.6777 & 0 & 5 & 1.6777 & 1.6777 & 0 \\
			2 & 5 & 8.0641 & 0 & 6.25 & 1.8141 & 1.8141 & 0 \\
			2 & 6 & 9.3949 & 0 & 7.5 & 1.8949 & 1.8949 & 0 \\
			2 & 7 & 10.6914 & 0 & 8.75 & 1.9414 & 1.9414 & 0 \\
			2 & 8 & 11.9678 & 0 & 10 & 1.9678 & 1.9678 & 0 \\
			2 & 9 & 13.2324 & 0 & 11.25 & 1.9824 & 1.9824 & 0 \\
			2 & 10 & 14.4905 & 0 & 12.5 & 1.9905 & 1.9905 & 0 \\
			2 & 11 & 15.7449 & 0 & 13.75 & 1.9949 & 1.9949 & 0 \\
			2 & 12 & 16.9972 & 0 & 15 & 1.9972 & 1.9972 & 0 \\
			\midrule
			3 & 1 & 2.9349 & 0 & 2.125 & 0.8099 & 0.8099 & 0 \\
			3 & 2 & 5.7994 & 0 & 4.25 & 1.5494 & 1.5494 & 0 \\
			3 & 3 & 8.4822 & 0 & 6.375 & 2.1072 & 2.1072 & 0 \\
			3 & 4 & 10.9777 & 0 & 8.5 & 2.4777 & 2.4777 & 0 \\
			3 & 5 & 13.3294 & 0 & 10.625 & 2.7044 & 2.7044 & 0 \\
			3 & 6 & 15.5863 & 0 & 12.75 & 2.8363 & 2.8363 & 0 \\
			3 & 7 & 17.7857 & 0 & 14.875 & 2.9107 & 2.9107 & 0 \\
			3 & 8 & 19.9517 & 0 & 17 & 2.9517 & 2.9517 & 0 \\
			\midrule
			4 & 1 & 3.9663 & 0 & 3.0625 & 0.9038 & 0.9038 & 0 \\
			4 & 2 & 7.8975 & 0 & 6.125 & 1.7725 & 1.7725 & 0 \\
			4 & 3 & 11.7534 & 0 & 9.1875 & 2.5659 & 2.5659 & 0 \\
			4 & 4 & 15.3994 & 0 & 12.25 & 3.1494 & 3.1494 & 0 \\
			4 & 5 & 18.8334 & 0 & 15.3125 & 3.5209 & 3.5209 & 0 \\
			4 & 6 & 22.1133 & 0 & 18.375 & 3.7383 & 3.7383 & 0 \\
		\end{tabular}
	\end{ruledtabular}
\end{table}

\subsection{\label{subsection:phase}Information-theoretic quantities for the phase estimation algorithm}
As mentioned in \ref{order_finding}, the procedure of phase estimation forms a key part of the order finding and factoring algorithms. Here we wish to study phase estimation as an oracle problem.

The standard formulation of the problem is as follows: suppose a unitary operator $U$ has an eigenvector $\ket{u}$ with eigenvalue $ e^{2\pi i \varphi} $ for $ \varphi \in \left[ 0, 1 \right) $. We have available an oracle that prepare the state $\ket{u}$, as well as oracles the perform controlled-$ U^{2^k} $ for any value of $k$, and we are tasked with finding $\varphi$ using as few queries as possible. The standard quantum algorithm for this problem uses a controlled-$ U^{2^k} $ gate, where the control qubit is set to $ \ket{+} $ and the target register is set to $\ket{u}$. This results in a phase of $ 2^k \varphi $ being ``kicked back'' to the control qubit. This procedure is performed for $ k = 0, \ldots, t-1 $ using $t$ separate control qubits, where $t$ depends on our desired precision. The final state of these qubits is (up to normalization) $ \sum_{l=0}^{2^t-1} e^{2\pi i \varphi l} \ket{l} $, and an inverse quantum Fourier transform is utilized to extract the phase.

It is not straightforward to cast this problem in our terms, since our formalism dictates that there should be only one family of oracles, and that the oracle identity is the only hidden information. In what follows, we suggest an alternative formulation of phase estimation that solves both of these issues, while retaining the essential features of the original problem and its quantum algorithm.

The idea is as follows: deviating from our usual assumption that $\mathcal{F}$ is a finite set, let it be the interval of possible phases, i.e. $ \mathcal{F} = \left[ 0, 1 \right) $; and suppose the random variable $F$ has the uniform distribution on this interval. For a fixed positive integer $n$, we want to define the problem of finding the first $n$ binary digits of the unknown phase $f$. I.e., $ f = 0. f_1 f_2 f_3 \ldots $ and we wish to find $ f_1, \ldots, f_n $. This is equivalent to finding the class of $f$ with respect to the partition $ \mathcal{F} = \bigsqcup_{j \in \mathcal{J}} A_j $, where $ \mathcal{J} = \left\{ 0, 1, 2, \ldots, 2^n-1 \right\} $ and $A_j \defeq \left\{ f \in \mathcal{F} \mid \frac{j}{2^n} \leq f < \frac{j+1}{2^n} \right\} $. Finally, we should describe the oracles $U_f$. These act on a register comprising $t$ qubits, and are diagonal in the computational basis with the following entries:
\begin{equation}
	U_f \ket{k} = e^{2\pi i f k} \ket{k} , \quad k \in \left\{ 0, 1, 2, \ldots, 2^t-1 \right\} ,
\end{equation}
where $ t \geq n $ depends on the amount of information we wish to extract.
Note that here we deviate from another one of our usual assumptions, i.e. that the oracle gates $U_f$ should only have entries $0$ or $1$ in the computational basis. We usually assume this to ensure that the oracle can be accessed by a classical algorithm; indeed, the problem of finding an unknown phase of a unitary gate does not directly apply to classical algorithms.

While this problem does not quite fall into the framework of the hidden subgroup problem, it can be solved with high probability using a single-query algorithm. This algorithm is given by $ V = H^{\otimes t} $ and $W$ the inverse quantum Fourier transform with respect to the group $ \mathbb{Z} / 2^t \mathbb{Z} $, i.e.
\begin{equation}
	\braket{k \vert W \vert l} = \frac{1}{2^{t/2}} e^{-\frac{2\pi i kl}{2^t} } , \quad k,l \in \left\{ 0, 1, 2, \ldots, 2^t-1 \right\} .
\end{equation}

Let us briefly present the information-theoretic quantities throughout this algorithm, which we find using a combination of analytic and numerical methods. The details appear in \Cref{app:phase_estimation}.

For the pre-query state, we find:
\begin{equation}
	C \left( \rho_Y^1 \right) = H \left( Y_1 \vert J \right) = H \left( Y_1 \right) = t 
\end{equation}
and
\begin{equation}
	S \left( \rho_Y^1 \right) = \chi_1 = I \left( J; Y_1 \right) = D_Y \left( \rho_{JY} ; Z^{\otimes t} \right) = 0 .
\end{equation}

After the query, we have $ H \left( Y_2 \vert J \right) = t $, $ C \left( \rho_Y^{2} \right) = 0 $ and $ H \left( Y_2 \right) = S \left( \rho_Y^{2} \right) = t $, implying $ I \left( J; Y_2 \right) = 0 $. $ \chi_2 $, which equals $D_Y^2 \left( \rho_{JY} ; Z^{\otimes n} \right) $, was found numerically. For sufficiently large difference $ t-n $, $ \chi_2 $ seems to approach $n$ (see $\chi$ in \ref{phase_estimation_simulation}), i.e. it is fairly close to the upper bound $ H \left( J \right) = n $.

Right before the final measurement, we have $ C \left( \rho_Y \right) = 0 $, $ H \left( Y \right) = S \left( \rho_Y \right) = t $ and $ \chi = \chi_2 $. Finding $ H \left( Y \vert J \right) $ analytically is challenging, but numerical simulations show that this value approaches $ t-n $ for large values of $t-n$. Hence, increasing the difference $t-n$ contributes to increasing the mutual information in two ways: it increases the Holevo quantity $\chi$ while decreasing the discord $ D_Y \left( \rho_{JY} ; Z^{\otimes n} \right) $. \ref{phase_estimation_simulation} depicts the information-theoretic quantities in the final step, for several values of $n$ and $t$.
\begin{table}
	\caption{\label{phase_estimation_simulation}Information-theoretic quantities for the final step of the phase estimation algorithm, found by simulating the algorithm's run over all possible inputs for $2 \leq n \leq 7$ and $n \leq t \leq 10 $. Notably, as $t$ increases $\chi$ approaches $n$ while the discord decreases ($D_Y$ is shorthand notation for $ D_Y \left( \rho_{JY} ; Z^{\otimes n} \right) $), resulting in the mutual information approaching $n$.
		Note that $ D_Y^2 = \chi_2 = \chi $, and all the other quantities for the first and second steps were found analytically, with the exact values given within the main body.}
	\tiny
	\begin{ruledtabular}
		\begin{tabular}{ c c c c c c c c c }
			$ n $ & $t$ & $ H \left( Y \right) $ & $ S \left( \rho_Y \right) $ & $ C \left( \rho_Y \right) $ &$ H \left( Y \vert J \right) $ & $ \chi $ & $ I \left( J ; Y \right) $ & $ D_Y $ \\ 
			\midrule
			2 & 2 & 2 &	2 & 0 & 1.3864 & 1.2090 & 0.6136 & 0.5954 \\
			2 & 3 & 3 &	3 & 0 & 1.9096 & 1.5250 & 1.0904 & 0.4346 \\
			2 & 4 & 4 &	4 & 0 & 2.5802 & 1.7220 & 1.4198 & 0.3022 \\
			2 & 5 & 5 &	5 & 0 & 3.3615 & 1.8405 & 1.6385 & 0.2019 \\
			2 & 6 & 6 &	6 & 0 & 4.2208 & 1.9099 & 1.7792 & 0.1306 \\
			2 & 7 & 7 &	7 & 0 & 5.1326 & 1.9498 & 1.8674 & 0.0823 \\
			2 & 8 & 8 &	8 & 0 & 6.0785 & 1.9723 & 1.9215 & 0.0507 \\
			2 & 9 & 9 &	9 & 0 & 7.0459 & 1.9848 & 1.9541 & 0.0307 \\
			2 & 10 & 10 & 10 & 0 & 8.0265 & 1.9918 & 1.9735 & 0.0183 \\
			\midrule
			3 & 3 & 3 &	3 & 0 & 1.5718 & 2.1865 & 1.4282 & 0.7583 \\
			3 & 4 & 4 &	4 & 0 & 2.0077 & 2.5146 & 1.9923 & 0.5223 \\
			3 & 5 & 5 &	5 & 0 & 2.6317 & 2.7170 & 2.3683 & 0.3487 \\
			3 & 6 & 6 &	6 & 0 & 3.3883 & 2.8380 & 2.6117 & 0.2263 \\
			3 & 7 & 7 &	7 & 0 & 4.2347 & 2.9087 & 2.7653 & 0.1434 \\
			3 & 8 & 8 &	8 & 0 & 5.1398 & 2.9492 & 2.8602 & 0.0890 \\
			3 & 9 & 9 &	9 & 0 & 6.0822 & 2.9720 & 2.9178 & 0.0542 \\
			3 & 10 & 10 & 10 & 0 & 7.0478 & 2.9847 & 2.9522 & 0.0325 \\
			\midrule
			4 & 4 & 4 &	4 & 0 & 1.6631 & 3.1810 & 2.3369 & 0.8441 \\
			4 & 5 & 5 &	5 & 0 & 2.0548 & 3.5120 & 2.9452 & 0.5668 \\
			4 & 6 & 6 &	6 & 0 & 2.6558 & 3.7157 & 3.3442 & 0.3716 \\
			4 & 7 & 7 &	7 & 0 & 3.4007 & 3.8374 & 3.5993 & 0.2381 \\
			4 & 8 & 8 &	8 & 0 & 4.2410 & 3.9084 & 3.7590 & 0.1494 \\
			4 & 9 & 9 &	9 & 0 & 5.1430 & 3.9490 & 3.8570 & 0.0920 \\
			4 & 10 & 10 & 10 & 0 & 6.0838 & 3.9719 & 3.9162 & 0.0558 \\
			\midrule
			5 & 5 & 5 &	5 & 0 & 1.7085 & 4.1797 & 3.2915 & 0.8881 \\
			5 & 6 & 6 &	6 & 0 & 2.0778 & 4.5114 & 3.9222 & 0.5892 \\
			5 & 7 & 7 &	7 & 0 & 2.6675 & 4.7154 & 4.3325 & 0.3829 \\
			5 & 8 & 8 &	8 & 0 & 3.4066 & 4.8373 & 4.5934 & 0.2438 \\
			5 & 9 & 9 &	9 & 0 & 4.2440 & 4.9083 & 4.7560 & 0.1523 \\
			5 & 10 & 10 & 10 & 0 & 5.1445 & 4.9490 & 4.8555 & 0.0935 \\
			\midrule
			6 & 6 & 6 &	6 & 0 & 1.7311 & 5.1793 & 4.2689 & 0.9104 \\
			6 & 7 & 7 &	7 & 0 & 2.0892 & 5.5112 & 4.9108 & 0.6004 \\
			6 & 8 & 8 &	8 & 0 & 2.6732 & 5.7154 & 5.3268 & 0.3886 \\
			6 & 9 & 9 &	9 & 0 & 3.4094 & 5.8372 & 5.5906 & 0.2467 \\
			6 & 10 & 10 & 10 & 0 & 4.2454 & 5.9083 & 5.7546 & 0.1537 \\
			\midrule
			7 & 7 & 7 &	7 & 0 & 1.7424 & 6.1793 & 5.2576 & 0.9216 \\
			7 & 8 & 8 &	8 & 0 & 2.0948 & 6.5112 & 5.9052 & 0.6060 \\
			7 & 9 & 9 &	9 & 0 & 2.6760 & 6.7153 & 6.3240 & 0.3914 \\
			7 & 10 & 10 & 10 & 0 & 3.4109 & 6.8372 & 6.5891 & 0.2481 \\
		\end{tabular}
	\end{ruledtabular}
\end{table}

\section{\label{sec:discussion}Discussion}
In this work we analyzed an insightful connection between oracle problems and quantum communication. We found a relation between optimality of single-query quantum algorithms and the optimization part in the definition of quantum discord. The measurement basis that minimizes quantum discord is also optimal in terms of the algorithm's performance (when measured by mutual information). This link is revealed only when formally considering the oracle as a separate physical subsystem with its own Hilbert space. Thus, Holevo's communication task bridges between quantum nonlocality and quantum computation.
Curiously, we have also found a lower bound on the mutual information, equivalent to the non-negativity of the relative entropy of coherence. Thus, quantum informational quantities arise naturally in the study of quantum algorithms for black-box problems, when choosing mutual information as the figure of merit.

Our results apply not only for single-query algorithms, but also for non-adaptive multiple-query algorithms. This is immediate from the insight detailed in \ref{non-adaptive}. Indeed, to study a $t$-query non-adaptive algorithm, one needs only to substitute the oracle unitary $ U_f $ by its $t$-th tensor power everywhere.

This paper yields an intuitive, physical heuristic for an optimal post-query unitary in any non-adaptive algorithm for an oracle problem. We also supply a function \eqref{I_max} of the pre-query state, whose maximization is a necessary condition for optimality. 
However, the maximization here seems to be difficult to perform in practice. Nonetheless, as demonstrated by the successful application of our framework to Quantum Likelihood Estimation (discussed in \Cref{subsec:QLE}), this mutual-information perspective provides a highly effective, exact objective function for optimizing iterative, hybrid quantum--classical subroutines, where continuous accumulation of partial information is paramount.

To illustrate our results, we depicted the information-theoretic quantities for several known quantum algorithms. This includes a thorough examination of the hidden subgroup problem for finite abelian groups. These examples may be used to develop intuition regarding the structure of successful algorithms. Let us elaborate on what these examples all have in common.
First, the mutual information always vanishes in the post-query phase. The measurement outcome entropy $ H \left( Y \right) $ remains high throughout all three stages. The conditional entropy $ H \left( Y \vert J \right) $ seems to follow a pattern of being initially high, remaining high in the intermediate step, and then reducing by the final value of the mutual information, $ I \left( J; Y \right) $. The Holevo quantity $\chi$ follows a somewhat contrasting pattern: it starts out as zero, obtaining the value $ I \left( J; Y \right) $ in the post-query phase, and then retaining this value in the final phase. The discord, however, is non-monotonous: it is initially zero, rises to a high value after the query, and then reduces to a low value in the final step. This post-query gap, $\chi - I(J;Y) = D_Y$, provides a clear, stage-by-stage mechanical interpretation of the algorithm: the oracle query ``stores'' information in the quantum state (reflected by a high Holevo quantity $\chi$ and high discord, while accessible information remains low). The final unitary then serves to ``unlock'' this stored information, minimizing the discord and allowing the extracted mutual information to reach its upper bound.
The coherence $ C \left( \rho_Y \right) $ always starts out with high value and ends up being zero in the final step. However, its post-query may be either low or high; in fact, it seems that high post-query coherence is the exception (that is the case in the Bernstein--Vazirani algorithm).

Furthermore, our comparative analysis of these algorithms reveals deeper structural connections regarding deterministic success. For instance, the Bernstein--Vazirani problem, as well as the Deutsch--Jozsa problem for small input sizes ($k=1,2$), can be viewed as specific instances of the Hidden Subgroup Problem where the hidden subgroup $H$ is promised to have either index $1$ or $2$, and the oracle function's codomain $X$ has cardinality $2$. In these specific cases, the existing algorithms utilize these algebraic promises to solve the problem with probability one using a single query. Motivated by these observations, a subsequent study~\cite{teeni2025one-query} introduces the general \emph{index-$q$ hidden subgroup problem}. In that follow-up work, we derived exact one-query algorithms that decide between index $1$ and $q$ for arbitrary abelian codomain structures, and exactly identify the hidden subgroup under explicit cyclicity and compatibility conditions, further illustrating the profound role of algebraic promises in one-query solvability.

Future work could seek to tackle broader questions. It would be interesting to identify conditions for an oracle problem to admit a quantum algorithm that succeeds with probability $1$. Alternatively, one may seek conditions for a problem to admit a quantum algorithm that outperforms any classical algorithm (i.e. a problem with quantum advantage).

While our framework successfully characterizes non-adaptive algorithms, fully adaptive algorithms---specifically ``quantumly adaptive'' protocols like Grover's algorithm, where multiple queries are applied sequentially with intermediate unitaries and no intermediate measurements---are currently outside its scope. Tracking the exact information-theoretic evolution of the state across multiple coherent, sequential channels presents a significant mathematical challenge.

However, our framework provides a conceptual roadmap for how this might be achieved in principle. An adaptive $t$-query algorithm can be modeled as an alternating sequence of operations, $T_f = W_t U_f W_{t-1} U_f \dots W_1 U_f V$. The classical-quantum state $\rho_{JFY}$ would continue to track the information flow across these stages, but the optimization landscape expands from a single pre- and post-query pair to a global, multi-stage parameter space. Within this lens, the intermediate unitaries $\left\{ W_1, \dots, W_{t-1} \right\}$ must serve a dual, coherent purpose: they must steer the ensemble to maximize the Holevo information ($\chi$) injected by subsequent queries, while preserving a geometric structure that allows the final unitary $W_t$ to successfully minimize the quantum discord ($D_Y$).
Formulating this sequential interaction as an optimal control or dynamic programming problem remains a highly interesting direction for future work, potentially offering a systematic method for discovering novel adaptive algorithms.

Finally, let us note that optimizing over the mutual information is not equivalent to optimizing over the probability of success (which is arguably a more natural performance measure). It is conceivable that studying the latter may lead to analogous insights and results. It may also be interesting to try adapting our framework and using it to study alternative models of quantum computation, e.g. adiabatic, one-way or topological quantum computers.

%

\appendix
\section{Remarks regarding the problem formulation}\label{app:formulation}
We list several further remarks regarding our problem formulation:
\begin{enumerate}
	\item We usually assume that $ \mathcal{F} $ is finite. Even if $ \mathcal{F} $ is infinite, we should make sure that $ \mathcal{J} $ is finite.
	
	\item For $F$ to be a well-defined random variable, it must obtain its values in $\mathbb{R}$. However, this will not pose a problem for us, as we will not be interested in probability-theoretic properties of $F$ such as expectation values, variance, correlations etc. In fact, we will only concern ourselves with information-theoretic quantities regarding $F$, namely its entropy and mutual information with other random variables. Thus, although we may assign real ``values'' to the elements of $\mathcal{F}$, the details of the specific map $ \mathcal{F} \rightarrow \mathbb{R} $ do not affect our analysis, as long as it is injective.
	
	\item Since we wish to compare the performances of classical and quantum algorithms, we must specify a classical protocol for oracle queries as well. Moreover, for the comparison to make any sense, there must be some kind of correspondence between the classical and quantum protocols. The standard way of achieving this, which we follow as well, is by imposing an additional constraint on the quantum protocol: $ \left\{ U_f \right\} \subseteq \mathrm{S}_M $; i.e., that the oracle-query gates $U_f$ consist only of $M \times M$ permutation matrices for some fixed $M$. This requirement ensures that oracle queries can only transform computational basis states into computational basis states. Thus, at least in the computational basis, an oracle query only computes some function of bit-strings (of length $M$). Such an operation admits an obvious translation into the language of classical computing. By ``padding'' if necessary, we can always assume that $ M = 2^m $ for some integer $m$, such that the unitary gates $U_f$ act on $m$ qubits. Note that the oracle itself is assumed to be classical.
	
	\item The oracle-query protocol in the $DJ(k)$ problem is even more restricted, as it leaves a subset of its input qubits unchanged in the computational basis (indeed, it only changes the value of a single qubit). Note that for $DJ(k)$ we have $ m = k+1 $.
	
	\item In most cases, we assume that the elements $f \in \mathcal{F}$ are in fact functions, and the oracle query computes the function for a given input. Therefore, we are really studying generalized Deutsch--Jozsa problems. Note that this setting is quite general.
\end{enumerate}

\section{Probability of success}\label{app:prob}
Recall that $ \ket{\psi_{\mathrm{fin}} \left( f \right)} = T_f \ket{ \psi_0 } $ is the final state of the quantum computer right before the measurement, where $\ket{ \psi_0 }$ is some fixed initial state in the computational basis. $T_f = W U_f V $ is the unitary that describes the entire quantum circuit. For the sake of simplicity, we assume that $\ket{ \psi_0 } = \ket{0 \ldots 0}$.
The possible values of $ Y $ are all of the sequences in $ \left\{ \pm 1 \right\}^n $. For convenience, we identify a $\pm 1$-sequence with the corresponding eigenstate, which is a computational basis state - i.e., a state of the form $ \ket{ y } $, where $ y \in \left\{ 0,1 \right\}^n $. Then, the probability to measure the outcome state $ \ket{y} $ given the oracle $f$ is:
\begin{equation}\label{Pr_Y_cond_F}
	\Pr \left( Y = y \mid F = f \right) = \abs{ \braket{ y \vert \psi_{\mathrm{fin}} \left( f \right) } }^2 = \abs{ \braket{ y \vert T_f \vert \psi_{0} } }^2 .
\end{equation}
Note that $ \ket{\psi_{\mathrm{fin}} \left( f \right)} $ should be the state right before the measurement; however, if some of the qubits are not measured, one should take the partial trace of $ \ket{\psi_{\mathrm{fin}} \left( f \right)} \bra{ \psi_{\mathrm{fin}} \left( f \right) } $ in order to discard the unmeasured qubits, or equivalently, compute the probabilities as
\begin{equation*}
	\Pr \left( Y = y \mid F = f \right) = \bra{ \psi_{\mathrm{fin}} \left( f \right) } \left( \ket{y} \bra{y} \otimes \mathbb{I}^{\otimes l} \right) \ket{ \psi_{\mathrm{fin}} \left( f \right) }
\end{equation*}
where $l$ is the number of discarded qubits. We shall mostly ignore these considerations from now on and simply assume all qubits are measured. This just means that $Y$ contains some extra information which is discarded.

It is convenient to compute $ \Pr \left( Y = y \right) $ using the law of total probability:
\begin{align}\label{Pr_Y}
	\Pr \left( Y = y \right) & = \EX_{f \in \mathcal{F}} \left[ \Pr \left( y \mid f \right) \right] = \EX_{f \in \mathcal{F}} \left[ \braket{ y \vert \psi_{\mathrm{fin}} \left( f \right) } \braket{ \psi_{\mathrm{fin}} \left( f \right) \vert y } \right] = \nonumber\\
	& = \bra{y} \EX_{f \in \mathcal{F}} \left[ \ket{ \psi_{\mathrm{fin}} \left( f \right) } \bra{ \psi_{\mathrm{fin}} \left( f \right) } \right] \ket{y} = \braket{ y \vert \rho_Y \vert y} ,
\end{align}
where $\EX$ denotes the expected value, $ \Pr \left( y \mid f \right) \defeq \Pr \left( Y=y \mid F=f \right) $, and:
\begin{equation}\label{rho_Y_def_supp}
	\rho_Y \defeq \EX_{f \in \mathcal{F}} \left[ \ket{ \psi_{\mathrm{fin}} \left( f \right) } \bra{ \psi_{\mathrm{fin}} \left( f \right) } \right] = \sum_{f \in \mathcal{F}} p_f T_f \ket{ \psi_0 } \bra{ \psi_0 } T_f^\dagger .
\end{equation}

To find probabilities related to the random variable $J$, let us rewrite \eqref{rho_Y_def_supp} as:
\begin{equation}\label{rho_Y_def_supp}
	\rho_Y = \sum_{j \in \mathcal{J}} p_j \sigma_j \; , \quad \sigma_j \defeq \frac{1}{p_j} \sum_{f \in A_j} p_f T_f \ket{ \psi_0 } \bra{ \psi_0 } T_f^\dagger ,
\end{equation}
where $ p_j \defeq \sum_{f \in A_j} p_f $. Using these notations, we have:
\begin{equation}
	\Pr \left( Y = y \mid J =j \right) = \braket{y \vert \sigma_j \vert y} = \frac{1}{p_j} \sum_{f \in A_j} p_f \abs{ \braket{ y \vert T_f \vert \psi_{0} } }^2 .
\end{equation}

Now, let $ \hat{J} $ denote the algorithm's guess for the value of $J$ (i.e. the ``estimator'' for $J$). Let us decide upon a scheme for how the algorithm chooses the value of $ \hat{J} $. For any fixed unitaries $V,W$ (or any fixed classical algorithm for producing the random variable $Y$), we can compute the conditional probabilities $ \Pr \left( J = j \mid Y = y \right) $; one way of doing so uses $p_j, \Pr \left( Y = y \mid J =j \right), \Pr \left( Y = y \right) $ and Bayes' theorem. Now, for each value of $ y \in \left\{ 0,1 \right\}^n $, we choose (apriori) a corresponding ``best guess'' $ g \left( y \right) $ for $J$:
\begin{equation}
	g \left( y \right) = \arg \max_{j \in \mathcal{J}} \Pr \left( J = j \mid Y = y \right) .
\end{equation}
(If there is more than one $ \arg \max $, we choose one of them arbitrarily.) Then, if we obtain $ Y=y $ in the final stage of the algorithm, we choose $ \hat{J} = g \left( y \right) $.

Assuming this scheme, we may now compute the average probability of success. Using the law of total probability:
\begin{equation}
	\bar{p}_s \defeq \Pr \left( \hat{J} = J \right) = \sum_{j \in \mathcal{J}} p_j \Pr \left( \hat{J} = j \mid J = j \right)
\end{equation}
To find $ \Pr \left( \hat{J} = j \mid J = j \right) $ we apply a conditional version of the law of total probability:
\begin{equation}
	\Pr \left( \hat{J} = j \mid J = j \right) = \sum_{y \in \left\{ 0,1 \right\}^n} \Pr \left( \hat{J} = j \mid J = j, Y = y \right) \cdot \Pr \left( Y = y \mid J = j \right)
\end{equation}
Since $ \hat{J} = g \left( y \right) $ for any given fixed value of $y$, we have:
\begin{equation}
	\Pr \left( \hat{J} = j \mid J = j, Y = y \right) = \begin{cases}
		1 ;& j = g \left( y \right) \\
		0 ;& j \neq g \left( y \right)
	\end{cases}
\end{equation}
i.e. $ \Pr \left( \hat{J} = j \mid J = j, Y = y \right) = \delta_{j, g \left( y \right)} $. Thus:
\begin{align}
	\bar{p}_s & = \sum_{j \in \mathcal{J}} \sum_{y \in \left\{ 0,1 \right\}^n} p_j \delta_{j, g \left( y \right)} \Pr \left( Y = y \mid J = j \right) = \sum_{y \in \left\{ 0,1 \right\}^n} p_{g \left( y \right)} \Pr \left( Y = y \mid J = g \left( y \right) \right) = \nonumber\\
	& = \sum_{y \in \left\{ 0,1 \right\}^n} \sum_{f \in A_{g \left( y \right)}} p_f \abs{ \braket{ y \vert T_f \vert \psi_{0} } }^2 = \sum_{y \in \left\{ 0,1 \right\}^n} p_{g \left( y \right)} \braket{y \vert \sigma_{g \left( y \right)} \vert y} .
\end{align}
Alternatively, we may wish to have the probability of success written as a sum over values of $j$:
\begin{equation}
	\bar{p}_s = \sum_{j \in \mathcal{J}} p_j \sum_{y \, : \, g \left( y \right) = j} \Pr \left( Y = y \mid J = j \right) = \sum_{j \in \mathcal{J}} p_j \sum_{y \, : \, g \left( y \right) = j} \braket{y \vert \sigma_{j} \vert y} .
\end{equation}
Let us take a closer look on the structure we have here. The function $ g \left( y \right) $ induces a decomposition of the Hilbert space of the quantum computer (i.e. the $n$-qubit space) into a direct sum of orthogonal subspaces:
\begin{equation}\label{H_C_decomposition}
	\mathcal{H}_C = \bigoplus_{j \in \mathcal{J}} \mathcal{H}_C^j , \quad \mathcal{H}_C^j \defeq \mathrm{span}_{\mathbb{C}} \left\{ \ket{y} \mid g \left( y \right) = j \right\} ,
\end{equation}
i.e., we group together all computational basis states $ \ket{y} $ such that $ g \left( y \right) = j $. Then:
\begin{equation}
	\bar{p}_s = \sum_{j \in \mathcal{J}} p_j \Tr{ \sigma_j \vert_{\mathcal{H}_C^j} } ,
\end{equation}
where $ \sigma_j \vert_{\mathcal{H}_C^j} $ is the restriction of $ \sigma_j $ to the subspace $ \mathcal{H}_C^j $. As an immediate corollary, we note that the probability of success is $ 1 $ iff $ \forall j \in \mathcal{J} , \, \Tr{ \sigma_j \vert_{\mathcal{H}_C^j} } = 1 $, i.e., every $ \sigma_j $ has support contained in the subspace $ \mathcal{H}_C^j $.

Define the conditional probability of success:
\begin{align}
	p_{s \vert f} & \defeq \Pr \left( \hat{J} = J \mid F = f \right) = \Pr \left[ \hat{J} = j \left( f \right) \mid F = f \right] = \nonumber\\ 
	& = \sum_{y \in \left\{ 0,1 \right\}^n} \Pr \left( \hat{J} = j \left( f \right) \mid F = f, Y = y \right) \cdot \Pr \left( Y = y \mid F = f \right) .
\end{align}
As before, we have: $ \Pr \left( \hat{J} = j \left( f \right) \mid F = f, Y = y \right) = \delta_{j \left( f \right), g \left( y \right)} $, hence:
\begin{equation}\label{p_s_f_quantum}
	p_{s \vert f} = \Pr \left[ g \left( Y \right) = j \left( f \right) \mid F = f \right] = \sum_{y \; : \; g \left( y \right) = j \left( f \right)} \Pr \left( Y = y \mid F = f \right) = \sum_{y \; : \; g \left( y \right) = j \left( f \right)} \abs{ \braket{ y \vert T_f \vert \psi_{0} } }^2 .
\end{equation}
Clearly we have $ \bar{p}_s = \sum_{f \in \mathcal{F}} p_f p_{s \vert f} $.

Moreover, we define the worst-case probability of success:
\begin{equation}\label{worst_case}
	p_{s,\min} \defeq \min_{f \in \mathcal{F}} p_{s \vert f} .
\end{equation}
Note that $ p_{s \vert f} $ and $ p_{s,\min} $ are independent of the probability distribution $ \left\{ p_f \right\} $.
Since $ \bar{p}_s $ is a convex combination of the $ p_{s \vert f} $, clearly $ \bar{p}_s \geq p_{s,\min} $ for \emph{any} choice of the probability distribution $ \left\{ p_f \right\} $. Furthermore, there exists a choice for the distribution $ \left\{ p_f \right\} $, such that we obtain an equality: pick $ f_\star \in \mathcal{F} $ for which $ p_{s \vert f} $ is minimal (i.e. $ \arg \min $ in the minimization \eqref{worst_case}), and take the deterministic distribution $ \delta_{f_\star} $ where $ p_{f_\star} = 1 $. Clearly, for this distribution we get $ \bar{p}_s = p_{s \vert f_\star } = p_{s,\min} $. Therefore, the worst-case probability of success equals the average probability of success with the distribution $ \left\{ p_f \right\} $ chosen \emph{adversarially}.

\subsection{Lower bound on probability of success}
Finally, let us derive a lower bound on $\bar{p}_s$ using mutual information. Start by expressing $\bar{p}_s$ with the posterior probabilities $ \Pr \left( J=j \mid Y=y \right) $:
\begin{align}
	\bar{p}_s & = \EX_{y} \left[ \Pr \left( \hat{J} = J \mid Y = y \right) \right] = \EX_{y} \left[ \Pr \left( J = g \left( y \right) \mid Y = y \right) \right] = \EX_{y} \left[ \max_{j \in \mathcal{J}} \Pr \left( J = j \mid Y = y \right) \right] = \nonumber\\
	& = \EX_{y} \left[ 2^{-H_\infty \left( J \mid Y = y \right) } \right] ,
\end{align}
where $ H_\infty \left( J \mid Y = y \right) = -\log \max_{j \in \mathcal{J}} \Pr \left( J = j \mid Y = y \right) $ is the \textit{min-entropy} of $J$ conditioned on $ Y=y $. Note that $2^{-x}$ is a convex function, thus
\begin{equation}\label{exponential_is_convex}
	\bar{p}_s = \EX_{y} \left[ 2^{-H_\infty \left( J \mid Y = y \right) } \right] \geq 2^{-\EX_{y} \left[ H_\infty \left( J \mid Y = y \right) \right] } = 2^{-H_\infty \left( J \mid Y \right)} .
\end{equation}

Since $ H_\infty \left( J \mid Y \right) \leq H \left( J \mid Y \right) $, we see that $ I \left( J; Y \right) \leq H \left( J \right) - H_\infty \left( J \mid Y \right) $, i.e.
\begin{equation}
	- H_\infty \left( J \mid Y \right) \geq I \left( J; Y \right) - H \left( J \right) .
\end{equation}
Combined with \eqref{exponential_is_convex}, we obtain $ \bar{p}_s \geq 2^{ I \left( J; Y \right) - H \left( J \right) } $. If $J$ has the uniform distribution, this implies $ \bar{p}_s \geq 2^{ I \left( J; Y \right) - \log \abs{\mathcal{J}} } $, as required.

\section{Relation between Holevo's bound and quantum discord}\label{app:discord}
In this section, we prove \eqref{Pawel_identity}.
Recall that the final state of the Hilbert space before measurement is:
\begin{equation}\label{final_state_supp}
	\rho_{JY} = \sum_{j \in \mathcal{J}} \sum_{f \in A_j} p_f \ket{j} \bra{j} \otimes T_f \ket{ \psi_0 } \bra{ \psi_0 } T_f^\dagger = \sum_{j \in \mathcal{J}} p_j \ket{j} \bra{j} \otimes \sigma_j .
\end{equation}
Clearly the reduced state of the $J$ subsystem is $ \rho_J = \sum_{j \in \mathcal{J}} p_j \ket{j} \bra{j} $. Since this is an eigenvalue decomposition of $ \rho_J $, it is clear that the von-Neumann entropy $ S \left( \rho_J \right) $ coincides with Shannon's entropy, $ H \left( J \right) $.
Moreover, due to the special classical-quantum form of \eqref{final_state_supp}, we can see that it is given in a block-diagonal form:
\begin{equation}
	\rho_{JY} = \begin{pmatrix}
		p_1 \sigma_1 & & \\
		& \ddots & \\
		& & p_{\abs{\mathcal{J}}} \sigma_{\abs{\mathcal{J}}}
	\end{pmatrix}
\end{equation}
That is, the eigenvalues of $ \rho_{JY} $ are $ p_j \lambda_k^j $, where $ \left\{ \lambda_k^j \right\}_{k=1}^{N} $ are the eigenvalues of $ \sigma_j $. The important thing to note here is the following implication:
\begin{align}
	S \left( \rho_{JY} \right) & = -\sum_{j \in \mathcal{J}} \sum_{k=1}^{N} p_j \lambda_k^j \log \left( p_j \lambda_k^j \right) = -\sum_{j \in \mathcal{J}} p_j \log p_j \sum_{k=1}^{N} \lambda_k^j -\sum_{j \in \mathcal{J}} p_j \sum_{k=1}^{N} \lambda_k^j \log \lambda_k^j = \nonumber\\
	& = H \left( J \right) + \sum_{j \in \mathcal{J}} p_j S \left( \sigma_j \right) .
\end{align}
Now, suppose the $Y$ subsystem is measured in the computational basis, and we obtain some outcome $ y \in \left\{ 0,1 \right\}^n $. This occurs with probability $ \Pr \left( Y = y \right) $, and the combined state then reduces to the following:
\begin{equation}
	\rho_{JY}^y = \sum_{j \in \mathcal{J}} \frac{ p_j \braket{y \vert \sigma_j \vert y} }{ \Pr \left( Y = y \right) } \ket{j} \bra{j} \otimes \ket{y} \bra{y} .
\end{equation}
Since $ \Pr \left( Y = y \vert J =j \right) = \braket{y \vert \sigma_j \vert y} $, we obtain using Bayes' theorem:
\begin{equation}
	\rho_{JY}^y = \sum_{j \in \mathcal{J}} \Pr \left( J = j \vert Y = y \right) \ket{j} \bra{j} \otimes \ket{y} \bra{y} .
\end{equation}
The oracle-class subsystem $J$ is in the state:
\begin{equation}
	\rho_{J}^{y} = \sum_{j \in \mathcal{J}} \Pr \left( J = j \vert Y = y \right) \ket{j} \bra{j} .
\end{equation}
Again, $ \left\{ \ket{j} \right\}_{j \in \mathcal{J}} $ are the eigenstates of $ \rho_{J}^{y} $, so we have an eigenvalue decomposition. Hence the eigenvalues of this measured state are $ \Pr \left( J = j \vert Y = y \right) $, and the conditional entropy is thus:
\begin{equation}
	S \left( \rho_J \vert \ket{y} \bra{y} \right) = -\sum_{j \in \mathcal{J}} \Pr \left( J = j \vert Y = y \right) \log \Pr \left( J = j \vert Y = y \right) = H \left( J \vert Y = y \right) .
\end{equation}
To find the conditional entropy with respect to the measurement (and not the measurement outcome), we must take the expected value:
\begin{align}\label{cond_entr}
	S \left( \rho_J \vert Z^{\otimes n} \right) & = \sum_{y \in \left\{ 0,1 \right\}^n} \Pr \left( y \right) S \left( \rho_J \vert \ket{y} \bra{y} \right) = \sum_{y \in \left\{ 0,1 \right\}^n} \Pr \left( y \right) H \left( J \vert Y = y \right) = H \left( J \vert Y \right) .
\end{align}
Substituting our expressions for $ S \left( \rho_{JY} \right) $ and $ S \left( \rho_J \vert Z^{\otimes n} \right) $ into the definition of non-optimized discord, we find:
\begin{align}\label{Pawel_identity_supp}
	D_Y \left( \rho_{JY} ; Z^{\otimes n} \right) & = S \left( \rho_Y \right) - H \left( J \right) -\sum_{j \in \mathcal{J}} p_j S \left( \sigma_j \right) + H \left( J \vert Y \right) = \nonumber\\
	& =	S \left( \rho_Y \right) -\sum_{j \in \mathcal{J}} p_j S \left( \sigma_j \right) - I \left( J; Y \right) ,
\end{align}
as required.

\section{Algorithms that succeed with probability $1$}\label{app:prob_1}
Recall that given the initial unitary gate $V$ we define the ensemble $ \left\{ \sigma_j^2 \right\} $ as follows:
\begin{equation}
	\sigma_j^2 = \sum_{f \in A_j} \frac{p_f}{p_j} U_f \ket{ \psi_1 } \bra{ \psi_1 } U_f^\dagger ,
\end{equation}
where $ \ket{ \psi_1 } = V \ket{ \psi_0 } $. Now, suppose there exists some $V$ such that $ \forall j \neq j' \in \mathcal{J} ,\; \Tr{\sigma_j^2 \sigma_{j'}^2} = 0 $. Using the results of \Cref{app:prob}, we see that the role of $W$ is to map the orthogonal support subspaces $ \mathcal{H}_C^{j,2} \defeq \left( \ker \sigma_j^2 \right)^\perp $ to subspaces that are spanned by computational basis vectors. This insight provides us with a simple way to construct $W$: first choose an orthonormal set $ \left\{ \ket{ \phi_{j,k} } \right\}_{j,k} $ in $ \mathcal{H}_C $ such that $ \left\{ \ket{ \phi_{j,k} } \right\}_k $ spans $ \mathcal{H}_C^{j,2} $; if necessary, complete it to form a basis of $ \mathcal{H}_C $ (we can assign the missing vectors to values of $j$ arbitrarily - this choice does not affect what follows). Equivalently, we may choose $ \left\{ \ket{ \phi_{j,k} } \right\}_{j,k} $ to be any orthonormal basis in which $ \sigma_j^2 $ are all simultaneously diagonal. Then, define $ W^\dagger $ such that it would map the computational basis to the basis $ \left\{ \ket{ \phi_{j,k} } \right\}_{j,k} $. Since $ W^\dagger $ maps one orthonormal basis to another, it is clearly unitary. Moreover, with this construction we have $ \sigma_{j} = W \sigma_j^2 W^\dagger $, and the probabilities have the following values:
\begin{equation}
	\Pr \left( Y = y \vert J = j \right) = \braket{y \vert \sigma_{j} \vert y} = \braket{y \vert W \sigma_j^2 W^\dagger \vert y} = \begin{cases}
		\braket{\phi_{j,k} \vert \sigma_j^2 \vert \phi_{j,k} } ; &  W^\dagger \ket{y} \in \mathcal{H}_C^{j,2} \\
		0 ; & W^\dagger \ket{y} \notin \mathcal{H}_C^{j,2}
	\end{cases}
\end{equation}
for some value of $k$. Furthermore:
\begin{equation}
	\Pr \left( Y = y \right) = \sum_{j \in \mathcal{J}} p_j \Pr \left( Y = y \vert J = j \right) = \sum_{j \in \mathcal{J}} p_j \braket{y \vert W \sigma_j^2 W^\dagger \vert y} ,
\end{equation}
where again, the summand corresponding to $j$ contributes to the sum iff $ W^\dagger \ket{y} \in \mathcal{H}_C^{j,2} $. Since the subspaces $ \mathcal{H}_C^{j,2} $ intersect only in the zero vector and span $ \mathcal{H}_C $ (if we redefine $ \mathcal{H}_C^{j,2} $ by adding the vectors from $ \ker \rho_Y $), there is exactly one term that contributes to the sum. This means (as expected) that for any value of $y$, if we measured the final state of the computer to be in the state $ \ket{y} $, then we know the value of $j$ with absolute certainty. Thus we have a function $ g \left( y \right) $ corresponding to the one defined in \Cref{app:prob}. Another way of seeing this is using Bayes' theorem and noting that the probabilities:
\begin{equation}
	\Pr \left( J = j \vert Y = y \right) = \frac{p_j \Pr \left( Y = y \vert J = j \right)}{\Pr \left( Y = y \right)}
\end{equation}
have only values $0$ and $1$.

Either way, with this definition of $ g \left( y \right) $ we have:
\begin{equation}
	\mathcal{H}_C^j = \mathrm{span}_{\mathbb{C}} \left\{ \ket{y} : W^\dagger \ket{y} \in \mathcal{H}_C^{j,2} \right\} = W \cdot \mathcal{H}_C^{j,2} .
\end{equation}
Thus, for each $j$, $ \sigma_j \vert_{\mathcal{H}_C^j} $ is the restriction of $ \sigma_j $ onto its support, and we obtain $ \Tr{\sigma_j \vert_{\mathcal{H}_C^j}} = 1 $ as expected.

\section{Lower bound on the mutual information}\label{app:lower_bound}
The outcome entropy $ H \left( Y \right) $ is computed by:
\begin{equation}
	H \left( Y \right) = -\sum_{y \in \left\{ 0,1 \right\}^n} \Pr \left( Y= y \right) \log \Pr \left( Y= y \right) .
\end{equation}
Therefore, by comparing with \eqref{Pr_Y} we may observe that if $ \rho_Y $ is diagonal in the computational basis, then $ H \left( Y \right) = S \left( \rho_Y \right) $. More generally, one may recall that the diagonal entries of a Hermitian matrix are majorized by its eigenvalues; i.e., $ \braket{y \vert \rho_Y \vert y} \prec \lambda \left( \rho_Y \right) $ (where both sides should be understood as vectors with $N = 2^n$ entries each). And since entropy is Schur-concave, we obtain:
\begin{equation}
	H \left( Y \right) \geq S \left( \rho_Y \right) .
\end{equation}

\section{Information-theoretic quantities for Bernstein--Vazirani}\label{app:BV}
Note that $\mathcal{F}$ is the set of the possible strings $a\in\left\{ 0,1\right\} ^{n}$, and we assume an equal probability for any $a\in\left\{ 0,1\right\} ^{n}$.

\subsection{Pre-query state}
Before the oracle query we have the state $ \ket{\psi_{b}} = \frac{1}{\sqrt{2^{n}}}\sum_{x=0}^{2^{n}-1} \ket{x} $, which is independent of the oracle identity $a \in \mathcal{F}$. Thus:
\begin{equation}
	\rho_Y^1 = \frac{1}{2^{n}} \sum_{a} \ket{ \psi_{b}\left(a\right) } \bra{\psi_{b}\left(a\right)} = \ket{ \psi_{b} } \bra{ \psi_{b} } .
\end{equation}
Now, for any $v\in\left\{ 0,1\right\} ^{n}$ we have $ \Pr\left(v|a\right)=\left|\left\langle v|\psi_{b}\right\rangle \right|^{2}= 2^{-n} $ and $ \Pr\left(v\right)=\left\langle v|\rho_{b}|v\right\rangle =\left|\left\langle v|\psi_{b}\right\rangle \right|^{2}=\Pr\left(v|a\right) $. Thus, the information-theoretic quantities are as follows:
\begin{equation}
	\begin{split}
		& H\left( Y \vert F \right)=H\left(Y\right)=-\sum_{v\in\left\{ 0,1\right\} ^{n}}\Pr\left(v\right)\log \Pr\left(v\right)=n \\
		& I\left( F; Y \right)=H \left( Y \right) - H \left( Y \vert F \right)=0 \\
		& S\left( \rho_Y^1 \right)= S \left( \ket{\psi_b} \bra{\psi_b} \right) = 0 \\
		& C \left( \rho_Y^1 \right) = H\left( Y \right) - S\left( \rho_Y^1 \right) = 0 .
	\end{split}
\end{equation}

\subsection{Intermediate state, after oracle}
The state right after the oracle query is:
\begin{equation}
	\ket{\psi_{m}\left(a\right)} = \frac{1}{\sqrt{2^{n}}} \sum_{x=0}^{2^{n}-1} \left(-1\right)^{f\left(x\right)} \ket{x} = \frac{1}{\sqrt{2^{n}}} \sum_{x=0}^{2^{n}-1} \left(-1\right)^{x\cdot a} \ket{x}.
\end{equation}
For any $ v \in \left\{ 0,1 \right\}^n $ we have:
\begin{equation}
	\left\langle v|\psi_{m}\left(a\right)\right\rangle =\frac{1}{\sqrt{2^{n}}}\sum_{x=0}^{2^{n}-1}\left(-1\right)^{x\cdot a}\left\langle v|x\right\rangle =\frac{1}{\sqrt{2^{n}}}\sum_{x=0}^{2^{n}-1}\left(-1\right)^{x\cdot a}\delta_{v,x}=\frac{1}{\sqrt{2^{n}}}\left(-1\right)^{v\cdot a}
\end{equation}
and
\begin{equation}
	\Pr\left(v|a\right)=\left|\left\langle v|\psi_{m}\left(a\right)\right\rangle \right|^{2}=\frac{1}{2^{n}}\left(-1\right)^{2v\cdot a}=\frac{1}{2^{n}} .
\end{equation}
Thus, we obtain:
\begin{equation}
	H\left(Y \vert F \right)  =-\frac{1}{\left|\mathcal{F}\right|} \sum_{a \in \mathcal{F}, v \in \left\{ 0,1\right\}^{n} } \Pr\left( v \vert a \right) \log \Pr\left( v \vert a \right)
	=-\frac{1}{2^{n}}2^{n}\cdot2^{n}\cdot\frac{1}{2^{n}}\cdot\left(-\log2^{n}\right)
	=n
\end{equation}
and
\begin{equation}
	\rho_Y^2 = \frac{1}{2^{n}}\sum_{a} \ket{\psi_{m}\left(a\right)} \bra{\psi_{m}\left(a\right)} = \frac{1}{2^{2n}} \sum_{a=0}^{2^{n}-1} \sum_{x=0}^{2^{n}-1} \sum_{y=0}^{2^{n}-1} \left(-1\right)^{a\left(x+y\right)} \ket{x} \bra{y} = \sum_{x,y = 0}^{2^n-1} c_{xy} \ket{x} \bra{y} ,
\end{equation}
where
\begin{equation}
	c_{xy} \defeq \frac{1}{2^{2n}}\sum_{a=0}^{2^{n}-1}\left(-1\right)^{a\left(x+y\right)} .
\end{equation}
Now, if $x+y=0 \, \mathrm{mod} \, 2$, i.e. $x=y$ then $c_{xy}=\frac{1}{2^{n}}$. If $x+y\neq0 \, \mathrm{mod}\, 2$ then the sum averages out to $0$ and we
get $c_{xy}=0$. In total we have $ c_{xy} = 2^{-n} \delta_{x,y} $, implying
\begin{equation}
	\rho_Y^2 = \frac{1}{2^{n}} \sum_{x=0}^{2^{n}-1} \ket{x} \bra{x} = \frac{\mathbb{I}}{2^{n}} ,
\end{equation}
i.e. $ \rho_Y^2 $ is fully mixed. Thus, we readily find that $ \rho_Y^2 $ has maximal entropy and zero coherence:
\begin{equation}
	S\left(\rho_Y^2\right) = n ,\; C \left( \rho_Y^2 \right) = 0.
\end{equation}
Moreover, for every  $ v \in \left\{ 0,1 \right\}^n $ we have $ \Pr\left(v\right) = \braket{v \vert \rho_Y^2 \vert v} = 2^{-n} $, hence
\begin{equation}
	H\left(Y\right)=-\sum_{v\in\left\{ 0,1\right\} ^{n}}\Pr\left(v\right)\log \Pr\left(v\right)=2^{n}\cdot\frac{1}{2^{n}}\log\left(2^{n}\right)=n .
\end{equation}
To conclude, right after the oracle query we have:
\begin{equation}
	\begin{split}
		& H\left(Y \vert F \right) =n\\
		& H\left(Y\right) =n\\
		& S\left(\rho_Y^2\right) =n\\
		& I\left( F; Y \right) =0 \\
		& C \left( \rho_Y^2 \right) = 0 .
	\end{split}
\end{equation}

\subsection{Final state}
The final state (without the ancilla) is:
\begin{equation}
	\ket{ \psi_{f}\left(a\right) } = \frac{1}{2^{n}} \sum_{y=0}^{2^{n}-1} \sum_{x=0}^{2^{n}-1} \left(-1\right)^{f\left(x\right)} \left(-1\right)^{x\cdot y} \ket{y} = \ket{a} .
\end{equation}
For a bit-string $v\in\left\{ 0,1\right\} ^{n}$ we have
\begin{equation}
	\Pr\left(v|a\right)=\left|\left\langle v|\psi_{f}\left(a\right)\right\rangle \right|^{2}=\left|\left\langle v|a\right\rangle \right|^{2}=\begin{cases}
		1 & v=a\\
		0 & v\neq a
	\end{cases} ,
\end{equation}
implying
\begin{equation}
	H\left(Y \vert F\right) =-\frac{1}{\left|\mathcal{F}\right|} \sum_{a\in\mathcal{F}, v\in\left\{ 0,1\right\} ^{n}} \Pr\left(v \vert a\right) \log \Pr\left(v\vert a\right)
	=-\frac{1}{2^{n}}2^{n}\left(\left(2^{n}-1\right)0\cdot\log0+1\cdot\log1\right)
	=0 .
\end{equation}
Now, 
\begin{equation}
	\rho_Y=\frac{1}{2^{n}}\sum_{a}|\psi_{f}\left(a\right)\rangle\langle\psi_{f}\left(a\right)|=\frac{1}{2^{n}}\sum_{a}|a\rangle\langle a|=\frac{1}{2^{n}}\mathbb{I} ,
\end{equation}
so as before we have:
\begin{equation}
	S\left( \rho_Y \right) = n, \; C \left( \rho_Y \right) = 0 .
\end{equation}
Moreover,
\begin{equation}
	\Pr\left(v\right) = \braket{v \vert \rho_Y \vert v} = \frac{1}{2^{n}} \braket{v \vert v} =\frac{1}{2^{n}} ,
\end{equation}
so
\begin{equation}
	H\left(Y\right)=-\sum_{v\in\left\{ 0,1\right\} ^{n}} \Pr\left(v\right) \log \Pr \left( v \right) = 2^{n} \cdot \frac{1}{2^{n}} \log \left(2^{n}\right) = n .
\end{equation}
Finally, we compute $ I\left(F;Y\right) = H\left( Y \right) - H\left( Y \vert F \right) = n $.
So in total we have
\begin{equation}
	\begin{split}
		& H\left(Y \vert F\right) =0\\
		& H\left(Y\right) =n\\
		& S\left(\rho_{f}\right) =n\\
		& I\left( F; Y \right) =n \\
		& C \left( \rho_Y^2 \right) = 0 .
	\end{split}
\end{equation}

\section{Information-theoretic quantities for the Shor--Kitaev algorithm}\label{app:SK}
The joint state of the oracle and computer at each step $i$ is of the form:
\begin{equation}
	\rho_{JFY}^i = \sum_{j \in \mathcal{J}} \sum_{f \in A_j} p_f \ket{j} \bra{j} \otimes \ket{f} \bra{f} \otimes \ket{ \psi_i \left( f \right) } \bra{\psi_i \left( f \right)} 
\end{equation}
where $ A_j $ is the set of all functions that hide the subgroup $ H_j $ of $G$ and $ \ket{ \psi_i \left( f \right) } $ is the state of the computer at step $i$. In particular, right before the query we have:
\begin{equation}
	\ket{\psi_1} = \frac{1}{\abs{G}^{1/2}} \sum_{g \in G} \ket{g} \ket{0} 
\end{equation}
After the query:
\begin{equation}
	\ket{\psi_2 \left( f \right)} = \frac{1}{\abs{G}^{1/2}} \sum_{g \in G} \ket{g} \ket{f \left( g \right)}
\end{equation}
Choosing a representative $ s \in gH_j $ for each coset $ gH_j \in G/H_j $, we can write:
\begin{equation}
	\ket{\psi_2 \left( f \right)} = \frac{1}{\abs{G}^{1/2}} \sum_{s \in G/H_j} \left( \sum_{h \in H_j} \ket{s+h} \right) \otimes \ket{f \left( s \right)}
\end{equation}
using the fact that $ f \left( g \right) $ is constant on cosets. The usual description of the algorithm states that here the second register should be measured and discarded (and the measurement outcome is not used). By the principle of deferred measurement, we can move this measurement to the end of the algorithm and obtain the exact same action of the circuit.

Finally, we perform a quantum Fourier transform on the first register. This has the following effect:
\begin{equation}
	\frac{1}{\abs{H_j}^{1/2}} \sum_{h \in H_j} \ket{s+h} \mapsto \sqrt{\frac{\abs{H_j}}{\abs{G}}} \sum_{\chi_g \in H_j^\perp} \chi_g \left( s \right) \ket{g}
\end{equation}
where $ \chi_g \in \hat{G} = \mathrm{Hom} \left( G, \mathrm{U} \left( 1 \right) \right) $ is a character of $G = \Pi_{i=1}^M \left( \mathbb{Z} / p_i  \mathbb{Z} \right)$ defined by
\begin{equation}
	\chi_g \left( l_1, \ldots l_M \right) \defeq \Pi_i e^{2\pi i l_i g_i/p_i} \quad \mathrm{for} \quad g = \left( g_1, \ldots, g_M \right) ;
\end{equation}
and $ H_j^\perp $ is the subgroup of $\hat{G}$ comprising the characters $\chi$ such that $ \mathrm{Res}_{H_j}^G \left( \chi \right) = \chi_e $, i.e. the characters that restrict to the trivial character on $H_j$.
So we have:
\begin{equation}
	\ket{\psi_{\mathrm{fin}} \left( f \right)} = \frac{\sqrt{\abs{H_j}}}{\abs{G}} \sum_{s \in G/H_j} \left( \sum_{\chi_g \in H_j^\perp} \chi_g \left( s \right) \ket{g} \right) \otimes \ket{f \left( s \right)} .
\end{equation}

Now we wish to analyze the information-theoretic quantities. The random variable $Y$ refers to the outcome of measuring only the first register in the computational basis (i.e. the one corresponding to elements of $G$). The pertinent matrices $ \rho_Y^i $ are:
\begin{equation}
	\rho_Y^i = \sum_{j \in \mathcal{J}} \sum_{f \in A_j} p_f \Tr_{O} \left( \ket{ \psi_i \left( f \right) } \bra{\psi_i \left( f \right)} \right)
\end{equation}
where $ \Tr_{O} $ denotes partial trace over the output register. Let us perform the computation for each of the three steps.
\subsection{Pre-query}
After discarding the output register, the state of the computer right before the query is a ``democratic superposition'', hence:
\begin{equation}
	C \left( \rho_Y^1 \right) = H \left( Y_1 \vert J \right) = H \left( Y_1 \right) = \log \abs{G}
\end{equation}
and of course, we have 
\begin{equation}
	S \left( \rho_Y^1 \right) = I \left( J; Y \right) = D_Y \left( \rho_{JY} ; Z^{\otimes n} \right) = 0 .
\end{equation}
This is explained in the paper as well.

\subsection{Post-query}
We have
\begin{equation}
	\rho_Y^2 = \sum_{j \in \mathcal{J}} p_j \rho_{Y}^{j,2}
\end{equation}
where
\begin{equation}
	\rho_{Y}^{j,2} = \frac{1}{\abs{G}} \sum_{s \in G/H_j} \sum_{h, h' \in H_j} \ket{s+h} \bra{s+h'} .
\end{equation}
Note:
\begin{equation}
	\rho_{Y}^{j,2} \ket{g} = \frac{1}{\abs{G}} \sum_{s \in G/H_j} \sum_{h, h' \in H_j} \ket{ s+h} \braket{s+h' \vert g} = \frac{1}{\abs{G}} \sum_{g' \in gH_j} \ket{g'}
\end{equation}
since any $g$ can be written uniquely as a sum $ g = s+h' $.
This matrix has the diagonal elements:
\begin{equation}
	\braket{ g \vert \rho_{Y}^{j,2} \vert g} = \frac{1}{\abs{G}} ,
\end{equation}
so $ H \left( Y_2 \vert J \right) = \log \abs{G} $;
and since $ \rho_Y^2 $ is a convex combination of matrices with these diagonal elements, it also has all diagonal elements equal $\frac{1}{\abs{G}}$, and we obtain $ H \left( Y_2 \right) = \log \abs{G} $ and $ I \left( J; Y_2 \right) = 0$. We would also later care about the inner products $\Tr \left( \rho_{Y}^{j',2} \rho_{Y}^{j,2} \right)$, so let us compute them here. The following
\begin{equation}
	\rho_{Y}^{j',2} \rho_{Y}^{j,2} \ket{g} = \frac{1}{\abs{G}} \sum_{\tilde{g} \in gH_j} \rho_{Y}^{j',2} \ket{\tilde{g}} = \frac{1}{\abs{G}^2} \sum_{\tilde{g} \in gH_j} \sum_{g' \in \tilde{g}H_{j'}} \ket{g'}
\end{equation}
implies
\begin{align}
	\braket{g \vert \rho_{Y}^{j',2} \rho_{Y}^{j,2} \vert g} & = \frac{1}{\abs{G}^2} \sum_{\tilde{g} \in gH_j} \mathbf{1}_{ H_{j'} } \left( g-\tilde{g} \right) = \nonumber\\
	& = \frac{1}{\abs{G}^2} \abs{ \left\{ \tilde{g} \in G \mid g-\tilde{g} \in H_j \cap H_{j'} \right\} } = \frac{1}{\abs{G}^2} \abs{ H_j \cap H_{j'} } ,
\end{align}
thus
\begin{equation}
	\Tr \left( \rho_{Y}^{j',2} \rho_{Y}^{j,2} \right) = \frac{\abs{ H_j \cap H_{j'} }}{\abs{G}} .
\end{equation}

Next, we should diagonalize $ \rho_Y^2 $. Start by writing down arbitrary matrix elements:
\begin{equation}
	\braket{g_r \vert \rho_{Y}^{j,2} \vert g_c} = \begin{cases}
		\frac{1}{\abs{G}} & g_r - g_c \in H_j \\
		0 & \mathrm{otherwise}
	\end{cases}
\end{equation}
so $ \rho_{Y}^{j,2} $ is a matrix with $\abs{G} / \abs{H_j}$ diagonal blocks, where each block is a $ \abs{H_j} \times \abs{H_j} $ square with all entries equal $1 / \abs{G}$. Each such block contributes a single nonzero eigenvalue $ \lambda = \abs{H_j} / \abs{G} $, so we obtain:
\begin{equation}
	S \left( \rho_{Y}^{j,2} \right) = \log \left( \abs{G} / \abs{H_j} \right) = \log \abs{G} -\log \abs{H_j} .
\end{equation}
We also have
\begin{equation}
	\braket{g_r \vert \rho_Y^2 \vert g_c} = \frac{1}{\abs{G}} \sum_{j \in \mathcal{J} \, : \, g_r-g_c \in H_j} p_j .
\end{equation}
If all probabilities $p_j$ are assumed to be equal, then
\begin{equation}
	\braket{g_r \vert \rho_Y^2 \vert g_c} = \frac{1}{\abs{G} \abs{\mathcal{J}}} \abs{ \left\{ j \in \mathcal{J} \mid g_r-g_c \in H_j \right\}} .
\end{equation}
Note that this matrix is real symmetric, and up to a constant all entries are non-negative integers. Moreover, we have $ \braket{g_r +a \vert \rho_Y^2 \vert g_c +a} = \braket{g_r \vert \rho_Y^2 \vert g_c} $ for all $a \in G$, so it is sufficient to describe the entries of the form $ \braket{g \vert \rho_Y^2 \vert e}= \frac{1}{\abs{G} } \Pr_J \left( g \in H_j \right) $. 

A simple observation greatly simplifies the computation. First, note that the ``input'' register is the Hilbert space $ \mathbb{C} \left[ G \right] $. This vector space carries the regular representation of $G$, defined via:
\begin{equation}
	g \cdot \ket{g'} \defeq \ket{g + g'} .
\end{equation}
This is a unitary representation, since
\begin{equation}
	\forall a,b \in G ,\quad \braket{a \vert g \cdot b} = \braket{a \vert g + b} = \delta_{a,g+b} = \delta_{a-g, b} = \braket{ a-g \vert b } = \braket{ -g \cdot a \vert b }
\end{equation}
implying $ g^\dagger = -g $. The second crucial observation is that $ \rho_Y^2 $, as a linear map in $\mathrm{End} \left( \mathbb{C} \left[ G \right] \right)$, commutes with this $G$-action, i.e. it is $G$-equivariant. Let us show this:
\begin{equation}
	\forall a,b \in G ,\quad \braket{a \vert \rho_Y^2 g \cdot \vert b} = \braket{a \vert \rho_Y^2 \vert g + b} = \braket{a -g \vert \rho_Y^2 \vert b} = \braket{ g^\dagger \cdot a \vert \rho_Y^2 \vert b} = \braket{a \vert g \cdot \rho_Y^2 \vert b} .
\end{equation}
Since $G$ is finite abelian, its regular representation reduces to a direct sum of all its irreps (each irrep being one-dimensional and appearing once), i.e. $ \mathbb{C} \left[ G \right] = \bigoplus_{\chi \in \hat{G}} V_\chi $, where $V_\chi$ is a one-dimensional space carrying the irrep $\chi$. By Schur's Lemma, for each $\chi$ we have that $\rho_Y^2 \vert_{V_\chi}$ acts as a scalar - that is, $ V_\chi $ are precisely the eigenspaces of $ \rho_Y^2 $. Moreover, since every $ \sigma_j^2 $ is also $G$-equivariant, we have that $ \left[ \sigma_j^2, \sigma{j'}^2 \right] = 0 $ for all $j, j'$; however, in general it is not true that $ \Tr \left( \sigma_j^{2} \rho_Y^{j',2} \right) =0 $ (otherwise a single query would have always sufficed to solve the HSP). Moreover, $ \left[ \sigma_j^{2}, \rho_Y^{2} \right] = 0 $ for every $ j \in \mathcal{J} $, implying that the minimized discord vanishes - $ D_Y \left( \rho_{JY}^2 \right) = 0 $.

We also have a straightforward description of the eigenvectors - they are given by the columns of the inverse quantum Fourier transform matrix, i.e. the complex conjugates of the rows of the character table of $G$:
\begin{equation}
	\ket{v_\chi} \defeq \frac{1}{\sqrt{\abs{G}}} \sum_{g \in G} \chi \left( g \right)^* \ket{g} .
\end{equation}
With that information, the eigenvalues of $ \rho_Y^2 $ are readily found:
\begin{multline}
	\lambda_\chi = \braket{ v_\chi \vert \rho_Y^2 \vert v_\chi} = \frac{1}{\abs{G}^2 } \sum_{j \in \mathcal{J}} p_j \sum_{g_r, g_c \in  G} \chi \left( g_r \right) \chi \left( g_c \right)^* \mathbf{1}_{H_j} \left( g_c-g_r \right) = \\
	= \frac{1}{\abs{G}^2 } \sum_{j \in \mathcal{J}} p_j \sum_{g_r, g_c \in  G} \chi \left( g_c -g_r \right)^* \mathbf{1}_{H_j} \left( g_c-g_r \right) =  \sum_{j \in \mathcal{J}} p_j \left( \frac{1}{\abs{G}} \sum_{g \in  G} \chi \left( g \right)^* \mathbf{1}_{H_j} \left( g \right) \right) .
\end{multline}
The expression in the parentheses equals:
\begin{equation}
	\frac{1}{\abs{G}} \sum_{g \in  G} \chi \left( g \right)^* \mathbf{1}_{H_j} \left( g \right) = \frac{1}{\abs{G}} \sum_{h \in H_j} \chi \left( h \right)^* \cdot 1 = \frac{\abs{H_j}}{\abs{G}} \braket{ \mathrm{Res}_{H_j}^G \left( \chi \right) , \chi_{\mathrm{triv}} }_{H_j} .
\end{equation}
Since $ \mathrm{Res}_{H_j}^G \left( \chi \right) $ is by itself a one-dimensional representation of $H_j$ (hence irreducible), it is orthogonal to the trivial irrep of $H_j$ iff it is not the trivial irrep. Thus we have:
\begin{equation}
	\lambda_\chi = \frac{1}{ \abs{G}} \sum_{j \in \mathcal{J} \, : \, \chi \in H_j^\perp} p_j \abs{H_j} .
\end{equation}
If $J$ is uniformly random, we obtain:
\begin{equation}
	\lambda_\chi = \frac{1}{\sqrt{\abs{G}} \abs{\mathcal{J}}} \braket{ c \vert v_\chi } ,
\end{equation}
where $ \ket{ c } = \sum_{g \in G} c_g \ket{g} $ and $c_g$ is defined to be the number of subgroups $H_j$ of which $g$ is an element. This is the most detailed we can get, as long as we keep our discussion completely general.

\subsection{Final state (pre-measurement)}
The Fourier transform maps the $\ket{v_\chi}$ basis to the computational basis, so right before the final measurement the matrix $ \rho_Y $ is diagonal. Thus, we have $ C \left( \rho_Y \right) = 0 $ and $ H \left( Y \right) = S \left( \rho_Y \right) = S \left( \rho_Y^2 \right) $. Moreover: 
\begin{equation}
	H \left( Y \vert J \right) = \sum_{j \in \mathcal{J}} p_j S \left( \sigma_j \right) = \log \abs{G} - \sum_{j \in \mathcal{J}} p_j \log \abs{H_j} .
\end{equation}
We also note that the minimized discord vanishes since $ \sigma_j $ pairwise commute; and the optimal measurement must be in the common eigenbasis, which is now the computational basis. Thus we have $ D_Y \left( \rho_{JY} ; Z^{\otimes n} \right) = 0 $, hence $ I \left( J; Y \right) = \chi \left( \left\{ \sigma_j ; p_j \right\} \right) $. 

Finally, we consider the quantities for $t$ queries. Since $t$ copies of the exact same circuit are performed in parallel, we simply have to replace $ \sigma_j $ by its $t$-th tensor power, $ \left( \sigma_j \right)^{\otimes t} $. Note that:
\begin{equation}
	\Tr \left[ \left( \sigma_j \right)^{\otimes t} \left( \rho_Y^{j'} \right)^{\otimes t} \right] = \Tr \left[ \left( \sigma_j \rho_Y^{j'} \right)^{\otimes t} \right] = \left[ \Tr \left( \sigma_j \rho_Y^{j'} \right) \right]^t = \left( \frac{\abs{ H_j \cap H_{j'} }}{\abs{G}} \right)^t
\end{equation}
so these matrices become ``more orthogonal'' with increasing values ot $t$, but they are never actually orthogonal. In particular, the $t$-th tensor powers are indexed by tuples $ \left( \chi_1, \ldots, \chi_t \right) $; and $ \left( \sigma_j \right)^{\otimes t} $ are all diagonal, with the following values:
\begin{equation}
	\left[ \diag \left( \sigma_j \right)^{\otimes t} \right] \left( \chi_1, \ldots, \chi_t \right) = \begin{cases}
		\left( \frac{\abs{ H_j }}{\abs{G}} \right)^t & \chi_1, \ldots, \chi_t \in H_j^\perp \\
		0 & \mathrm{otherwise}
	\end{cases}
\end{equation}
Recall that $Y_t$ denotes the outcome of measuring all $t$ input registers at the end of the algorithm. It is straightforward that
\begin{equation}
	H \left( Y_t \vert J=j \right) = S \left[ \left( \sigma_j \right)^{\otimes t} \right] = t \cdot H \left( Y \vert J=j \right) = t \left( \log \abs{G} -\log \abs{H_j} \right) ,
\end{equation}
hence $ H \left( Y_t \vert J \right) = t\cdot H \left( Y \vert J \right) = t \log \abs{G} -t \sum_{j \in \mathcal{J}} p_j \log \abs{H_j} $. Computing $H \left( Y_t \right)$ is a bit more complicated. It equals the von-Neumann entropy of the matrix $ \rho_Y^{(t)} = \sum_{j \in \mathcal{J}} p_j \left( \sigma_j \right)^{\otimes t} $, which has the following diagonal entries:
\begin{equation}
	\left[ \diag \rho_Y^{(t)} \right] \left( \chi_1, \ldots, \chi_t \right) = \sum_{j \in \mathcal{J} \, : \, \chi_1, \ldots, \chi_t \in H_j^\perp} p_j \left( \frac{\abs{ H_j }}{\abs{G}} \right)^t .
\end{equation}
$H \left( Y_t \right) $ is the entropy of these values, but in general there is no nice expression for it; the same holds for the mutual information $ I \left( J; Y_t \right) $.

\section{Information-theoretic quantities for Simon's algorithm}\label{app:Simon}
Let us apply the results of the previous section. While the pre-query quantities are readily obtained, for the post-query stage we need to find the eigenvalues of $ \rho_Y^2 $ using \Cref{subsub:HSP_post_query}. Recall that the character table of $ \left( \mathbb{Z} / 2 \mathbb{Z} \right)^n $ is given by $ \left( \sqrt{2} H \right)^{\otimes n} $ where $H$ is the Hadamard matrix, and $ \sqrt{2} H $ is the character table of $ \mathbb{Z} / 2 \mathbb{Z} $. A concise description for this matrix is as follows: let $ g, \chi \in \left\{ 0, 1 \right\}^n $ be elements of $G, \hat{G}$ respectively, both represented by strings of $n$ bits. Then $ \chi \left( g \right) = \left( -1 \right)^{\chi \cdot g} $, where $ \cdot $ is an inner product of bit strings. Thus, $ \chi \in H_s^\perp $ iff $ \chi \cdot s = 0 \; \mathrm{mod} \; 2 $, i.e. $ H_s^\perp $ comprises all bit strings that are orthogonal to $s$. We find:
\begin{equation}
	\lambda_\chi = \frac{1}{2^n} \left( p_0 + 2 \sum_{s \neq 0 \, : \, \chi \perp s} p_s \right) .
\end{equation}
Recall we are assuming the uniform distribution, i.e. $ p_s = \frac{1}{2^n} $ for all $s $. Noting that each $ \chi \neq 0 $ is perpendicular to exactly $ 2^{n-1} -1 $ nonzero strings. We find:, we obtain:
\begin{align}
	\forall \chi \neq 0 , \; \lambda_{\chi} = \frac{1}{4^n} \left( 1 + 2 \left( 2^{n-1} -1 \right) \right) = \frac{2^n-1}{4^n}  .
\end{align}
Hereon we denote this eigenvalue by $ \lambda_{\chi \neq 0} $. For $\chi = 0$ we have
\begin{equation}
	\lambda_0 = \frac{1}{2^n} \left( p_0 + 2 \sum_{s \neq 0} p_s \right) = \frac{1}{4^n} \left[ 1 + 2 \left( 2^n- 1 \right) \right] = \frac{2^{n+1} -1}{4^n} .
\end{equation}
Hence the entropy is
\begin{align}
	S \left( \rho_Y^2 \right) = \frac{2^{n+1} -1}{4^n} \left[ 2n- \log \left( 2^{n+1}-1 \right) \right]
	+ \frac{\left( 2^n-1 \right)^2}{4^n} \left[ 2n- \log \left( 2^n-1 \right) \right] \approx n 
\end{align}
up to an order of $2^{-n}$. We also note that $ H \left( J \right) = n $ and $ C \left( \rho_Y^2 \right) \approx 0 $ up to an order of $2^{-n}$. The discord is
$ D_Y \left( \rho_{JY}^2 ; G \right) \approx 1 $, up to the same order.

In the final stage we have $ C \left( \rho_Y \right) = 0 $, $ H \left( Y \right) = S \left( \rho_Y \right) \approx n $ and $ D_Y \left( \rho_{JY} ; G \right) = 0 $ as before; $ H \left( Y \vert J \right) \approx n-1$ and $ I \left( J; Y \right) = \chi \left( \left\{ \rho_{Y}^{j}; p_j \right\} \right) \approx 1 $. 

\section{Information-theoretic quantities for phase estimation}\label{app:phase_estimation}
In this section we compute the information-theoretic quantities for the phase estimation algorithm (with the slight modifications described in the main text). Following the notation and analysis in subsection 5.2.1 here~\cite{nielsen2002quantum}, let $ b \defeq \left\lfloor 2^t f \right\rfloor $ and $\delta \defeq f -b/2^t$. The post-query state is:
\begin{equation}\label{phase_estim_post_query}
	\ket{\psi_2 \left( f \right)} = \frac{1}{2^{t/2}} \sum_{k=0}^{2^t-1} e^{2\pi i fk} \ket{k} ,
\end{equation}
and the final state is:
\begin{equation}
	\ket{\psi_\mathrm{fin} \left( f \right)} = \frac{1}{2^t} \sum_{l=0}^{2^t-1} \frac{1- e^{2\pi i\left( 2^t \delta -l \right) } }{ 1- e^{2\pi i\left( \delta -l/2^t \right) }} \ket{\left( b+l \right) \mathrm{mod} \, 2^t} .
\end{equation}

Noting that the pre-query state is a democratic superposition, we have:
\begin{equation}
	C \left( \rho_Y^1 \right) = H \left( Y_1 \vert J \right) = H \left( Y_1 \right) = t 
\end{equation}
and
\begin{equation}
	S \left( \rho_Y^1 \right) = \chi_1 = I \left( J; Y_1 \right) = D_Y \left( \rho_{JY} ; Z^{\otimes t} \right) = 0 .
\end{equation}

Let us compute $ \sigma_j^{2} $. Recall we have defined this state as
\begin{equation*}
	\sigma_j^{2} = \frac{1}{p_j} \sum_{f \in A_j} p_f \ket{\psi_2 \left( f \right)} \bra{\psi_2 \left( f \right)} ,
\end{equation*}
where $ \ket{\psi_2 \left( f \right)} = U_f V \ket{\psi_0} $. However, in our case the sets $ A_j $ are countably infinite. There are two ways to tackle this issue. First, we can replace the sum by an integral, multiplied by a normalization constant that will be found later. The second approach is to treat the problem as if $f$ can be written using a finite number $s$ of binary digits, where $ s \geq t $, and ultimately take the limit $ s \rightarrow \infty $. As one should expect, these two approaches yield the same result, so we only show in detail the former, i.e. we write:
\begin{equation}\label{phase_estimation_propto}
	\sigma_j^{2} \propto \int_{f \in A_j} df \ket{\psi_2 \left( f \right)} \bra{\psi_2 \left( f \right)} = \frac{1}{2^{t}} \int_{j/2^n}^{ (j+1)/2^n } df \sum_{k,l=0}^{2^t-1} e^{2\pi i f \left( k-l \right)} \ket{k} \bra{l} .
\end{equation}
Let us compute the following integral for $k \neq l$:
\begin{align}
	& \int_{j/2^n}^{ (j+1)/2^n } e^{2\pi i f \left( k-l \right)} df = \left. \frac{e^{2\pi i f \left( k-l \right)}}{2\pi i \left( k-l \right)} \right\vert_{f = j/2^n}^{(j+1)/2^n} = \frac{ e^{ \frac{2\pi i \left( k-l \right) \left( j+1 \right)}{2^n} }- e^{\frac{2\pi i \left( k-l \right) j}{2^n} } }{2\pi i \left( k-l \right)} = \nonumber\\
	& = 
	\frac{ 1 }{ \pi \left( k-l \right)} \sin \left( \frac{\pi \left( k-l \right)}{2^n} \right) e^{\frac{2\pi i \left( k-l \right) }{2^n} \left( j+\frac{1}{2} \right) } = \frac{1}{2^n} \mathrm{sinc} \left( \frac{k-l}{2^n} \right) e^{\frac{2\pi i \left( k-l \right) }{2^n} \left( j+\frac{1}{2} \right) }
\end{align}
where $ \mathrm{sinc} \left( x \right) = \frac{\sin \left( \pi x \right)}{\pi x} $ is the normalized sinc function.
Note we get we fraction independent of $j$, multiplied by a $j$-dependent phase. For $ k=l $ the value of the above integral is $ 1 / 2^n $. Since the diagonal entries are all equal, to obtain a normalized state we must multiply the RHS of \eqref{phase_estimation_propto} by $2^n$.
Thus:
\begin{align}
	\sigma_j^{2} & = \frac{1}{2^t} \sum_{k=0}^{2^t-1} \ket{k} \bra{k} + \frac{1}{2^{t-n}} \sum_{k \neq l} \frac{e^{ 2\pi i \left( k-l \right) / 2^n } -1 }{2\pi i \left( k-l \right)} e^{\frac{2\pi i \left( k-l \right) j}{2^n} } \ket{k} \bra{l} = \nonumber\\
	& = \frac{1}{2^t} \sum_{k=0}^{2^t-1} \ket{k} \bra{k} +\frac{2^n}{2^{t}} \sum_{k \neq l} \frac{e^{2\pi i \left( k-l \right) \frac{j+1}{2^n} }- e^{2\pi i \left( k-l \right) \frac{j}{2^n} } }{ 2\pi i \left( k-l \right) } \ket{k} \bra{l} = \nonumber\\
	& = \frac{1}{2^t} \sum_{k=0}^{2^t-1} \ket{k} \bra{k} + \frac{1}{2^t} \sum_{k \neq l} \mathrm{sinc} \left( \frac{k-l}{2^n} \right) e^{\frac{2\pi i \left( k-l \right) }{2^n} \left( j+\frac{1}{2} \right) } \ket{k} \bra{l} .
\end{align}

Observing that the diagonal entries of $ \sigma_j^{2} $ all equal $ 2^{-t} $, we see that $ H \left( Y_2 \vert J \right) = t $.
Next, we compute $ \rho_Y^{2} $:
\begin{equation}
	\rho_Y^{2} = \frac{1}{2^t} \sum_{k=0}^{2^t-1} \ket{k} \bra{k} + \frac{1}{2^{n+t}} \sum_{k \neq l} \mathrm{sinc} \left( \frac{k-l}{2^n} \right) \ket{k} \bra{l} \sum_{j=0}^{2^n-1} e^{\frac{2\pi i \left( k-l \right) }{2^n} \left( j+\frac{1}{2} \right) } ,
\end{equation}
and for $k \neq l$ we have:
\begin{equation}
	\sum_{j=0}^{2^n-1} e^{\frac{2\pi i \left( k-l \right) }{2^n} \left( j+\frac{1}{2} \right) } = e^{\frac{2\pi i \left( k-l \right) }{2^{n+1}} } \frac{1- e^{2\pi i \left( k-l \right) } }{1- e^{2\pi i \left( k-l \right) /2^n } } = 0 ,
\end{equation}
hence $ \rho_Y^{2} $ is the fully mixed state, implying $ C \left( \rho_Y^{2} \right) = 0 $ and $ H \left( Y_2 \right) = S \left( \rho_Y^{2} \right) = t $. Thus we have $ I \left( J; Y_2 \right) = 0 $.

It is not straightforward to diagonalize the matrices $ \left\{ \sigma_j^{2} \right\} $. For nontrivial values of $n,t$ they do not commute, so $ \chi_2 < H \left( J \right) $. However, numerical simulations show that these matrices have the same eigenvalues. Moreover, $ S \left( \sigma_j^{2} \right) $ seems to approach $ t-n $ for sufficiently large difference $ t-n $ (see the table in the paper). This implies $ \chi_2 \approx n $.

Right before the final measurement, $ \rho_Y $ is still the fully mixed state, so we have $ C \left( \rho_Y \right) = 0 $ and $ H \left( Y \right) = S \left( \rho_Y \right) = t $. The values of $ S \left( \sigma_j \right) $ are the same as in the second step, hence so is $ \chi $. Simulation results suggest that $ H \left( Y \vert J \right) $ approaches $ t-n $ for large values of $t-n$. Note this also implies that the states $ \sigma_j $ tend to be incoherent (i.e. diagonal) in the Fourier-inverted basis, in the limit of high $ t-n $.

\end{document}